\documentclass[a4paper,11pt]{article}
\pdfoutput=1 % if your are submitting a pdflatex (i.e. if you have
\usepackage{jheppub,bm} % for details on the use of the package, please
\usepackage[T1]{fontenc} % if needed
\usepackage{slashed}
\def\MC@NLO{{\sc MC@NLO}}

\def\PYTHIA8{{\sc PYTHIA8}}
\def\HERWIG++{{\sc HERWIG++}}

\def\tauzerocut{{\tau}^{\mbox{\tiny{cut}}}}

\def\as{\alpha_s}

\def\spa#1.#2{\langle #1  #2\rangle}
\def\spab#1.#2.#3{\langle #1|#2|#3]}
\def\spb#1.#2{[ #1 #2]}
\DeclareMathOperator{\FFtme}{ {\rm F}^{2me}_{4F}}
\DeclareMathOperator{\FFom}{ {\rm F}^{1m}_{4F}}

\def\beq{\begin{equation}}
\def\eeq{\end{equation}}
\def\beqn{\begin{eqnarray}}
\def\eeqn{\end{eqnarray}}

\allowdisplaybreaks[4]

\begin{document}

\title{Associated production of a Higgs boson at NNLO}

%% %simple case: 2 authors, same institution
\author[a]{John M. Campbell}
\author[b]{R. Keith Ellis}
\author[c]{and Ciaran Williams}

\affiliation[a]{Fermi National Accelerator Laboratory,\\PO Box 500, Batavia, IL 60510, USA}
\affiliation[b]{Institute for Particle Physics Phenomenology,\\Department of Physics, Durham University, Durham DH1 3LE, United Kingdom}
\affiliation[c]{Department of Physics, University at Buffalo\\ The State University of New York, Buffalo 14260 USA}

% e-mail addresses: one for each author, in the same order as the authors
\emailAdd{johnmc@fnal.gov}
\emailAdd{keith.ellis@durham.ac.uk}
\emailAdd{ciaranwi@buffalo.edu}

\newcommand{\zero}{{(0)}}
\newcommand{\one}{{(1)}}
\newcommand{\two}{{(2)}}
\newcommand{\ztwo}{\zeta_2}
\newcommand{\zthree}{\zeta_3}
\newcommand{\cf}{C_F}
\newcommand{\ca}{C_A}
\newcommand{\nf}{n_f}
\newcommand{\cfs}{C_F^2}
\newcommand{\Tcm}{\tau_{cm}}

\abstract{ 
In this paper we present a Next-to-Next-to Leading Order
(NNLO) calculation of the production of a Higgs boson in association
with a massive vector boson. We include the decays of the unstable
Higgs and vector bosons, resulting in a fully flexible parton-level
Monte Carlo implementation. We also include all
$\mathcal{O}(\alpha_s^2)$ contributions that occur in production for
these processes: those mediated by the exchange of a single
off-shell vector boson in the $s$-channel, and those which arise
from the coupling of the Higgs boson to a closed loop of
fermions. We study final states of interest for Run II
phenomenology, namely $H\rightarrow b\overline{b}$, $\gamma\gamma$ and
$WW^*$. The treatment of the $H\rightarrow b\overline{b}$ decay
includes QCD corrections at NLO. We use the recently developed
$N$-jettiness regularization procedure, and study its viability in
the presence of a large final-state phase space by studying
$pp\rightarrow V(H\rightarrow WW^*) \rightarrow$ leptons.}

\preprint{
\noindent IPPP/15/78 \\
\hspace*{\fill} FERMILAB-PUB-16-001-T}
\maketitle
\flushbottom
\section{Introduction}

% Introduction for VH paper 

% General Higgs blurb 

Run II of the LHC promises to shed new light on the mysteries behind
the breaking of the electroweak symmetry. The standout result from Run
I of the LHC was the discovery of a Higgs
boson~\cite{Aad:2012tfa,Chatrchyan:2012xdj}. One of the principal
physics goals of Run II is to pin down the precise nature of the Higgs
boson and in particular how it interacts with the other 
particles of the Standard
Model (SM). In order to do this a range of Higgs production and decay
processes must be studied in greater detail than ever before. A
significant improvement that is expected in Run II analyses
is their ability to study the Higgs differentially for a
wider range of processes.

One such fascinating process is the production of a Higgs boson in
association with a $W$ or $Z$ electroweak vector boson,
i.e. $pp\rightarrow VH$ where $V$ denotes the vector boson. At LHC energies these processes are the third
($V=W$) and fourth ($V=Z$) largest production channels. $VH$ production
is somewhat special, in that it proceeds at Leading Order (LO) through
an $s$-channel Feynman diagram. This results in the opportunity to
probe the $VVH$ vertex at high momentum transfer while keeping the
final state vector and Higgs bosons on-shell, for instance by looking
in the region of large $m_{VH}$. This is an interesting region to
study, since contributions arising from physics beyond the Standard Model (BSM) 
may induce a momentum dependent term in the $VVH$ vertex~\cite{Buchmuller:1985jz,Hagiwara:1993ck,Contino:2013kra}. 
New physics at TeV scales would modify the SM cross-section
at the level of a few percent.  Accordingly, it is essential that the SM
cross-section is known at this level or better.

A second distinguishing feature of the $pp\rightarrow VH$ process is
the ability to study the decay of the Higgs boson to a pair of bottom
quarks, $H \to b\overline{b}$. Such a decay is extremely
difficult to measure in inclusive Higgs boson production,
given the small rate of $gg\rightarrow H\rightarrow b\overline{b}$
compared to the QCD production of the same final state. It is essential
that the decay $H\rightarrow b\overline{b}$ is measured experimentally
since it provides a direct measurement of the coupling of the Higgs boson to
fermions.  Moreover,  since it dominates the total width, the uncertainty on
this branching ratio feeds into other searches, for instance,
measurements of the Higgs invisible branching ratio. The presence of
the vector boson in the final state allows experimental analyses to
have manageable backgrounds, in particular when the Higgs is highly
boosted and the two $b$-quarks reside inside a fat
jet~\cite{Butterworth:2008iy}. Again it is essential that accurate
theoretical predictions are available, with the ability to apply
intricate final state phase space selection requirements.

% Discussion of VH in literature 

Given its importance, the $pp\rightarrow VH$ processes have been
extensively studied in the theoretical literature. At LO the topology
is essentially the same as that of Drell-Yan (DY) production, and this
was utilized to obtain the first Next-to-Next-to Leading Order (NNLO)
predictions for on-shell bosons in ref.~\cite{Brein:2003wg}. However at
$\mathcal{O}(\alpha_s^2)$ a second type of diagram appears, in which
instead of coupling to the vector boson, the Higgs is radiated from a
closed loop of heavy fermions. These ``$y_t$'' pieces\footnote{We refer to
  these pieces with the label $y_t$ in this paper, despite the fact that we also include the
  $gg\rightarrow Z^*\rightarrow ZH$ contributions (that do not go like
  $y_t$) in this term.}  were computed for on-shell vector
bosons in ref.~\cite{Brein:2011vx}. A fully differential calculation,
including the decays of the bosons, was presented for the DY parts (i.e. neglecting the $y_t$ terms) of
$WH$ in refs.~\cite{Ferrera:2011bk,Ferrera:2013yga} and of $ZH$ in
ref.~\cite{Ferrera:2014lca}. A subset of the $y_t$ diagrams,
corresponding to those which are initiated by a pair of gluons
$gg\rightarrow HZ$, was also included in the calculation of
ref.~\cite{Ferrera:2014lca}. 
%v2 CW
A primary motivation of this paper is to extend the calculations of refs.~\cite{Ferrera:2011bk,Ferrera:2013yga,Ferrera:2014lca}
to fully account for the contributions discussed in ref.~\cite{Brein:2011vx} in a flexible Monte Carlo code. We will also 
extend the range of Higgs boson decays beyond the two-body ones presented previously.
Electroweak corrections were calculated
in ref.~\cite{Ciccolini:2003jy,Denner:2011id} while resummation effects have
been studied in refs.~\cite{Dawson:2012gs,Shao:2013uba,Li:2014ria}. There has also been
significant progress in matching fixed order calculations to parton
shower Monte Carlos, allowing for full event simulation. An
implementation of the $VH$ process in the POWHEG formalism was
presented in~\cite{Hamilton:2009za} and extended to merge with the
$VH+$jet process in ref.~\cite{Luisoni:2013cuh}. A SHERPA implementation that
merges the $VH$ and $VH+$jet processes was also presented recently in
ref.~\cite{Goncalves:2015mfa}.

% Experimental studies 

% Discussion of SCET/NNLO 

The historical bottleneck for NNLO computations was in the
construction of regularization schemes to handle the InfraRed (IR)
singularities. These singularities are ubiquitous in a NNLO
calculation, since they occur in the two-loop (double virtual),
one-loop~$\times$~real (real-virtual) and the real-real part of the
calculation. The situation is made more complicated by the different
dimensionality of the phase space in each part. The double-virtual has
the same dimension as the Born, and IR singularities manifest
themselves as poles in an $\epsilon$ expansion (where $d=4-2\epsilon$,
with $\epsilon$ parameterizing excursions from four dimensions). The
real-virtual has one additional parton in the final state, and
possesses IR singularities which manifest themselves as $\epsilon$
poles, and when the emitted parton becomes unresolved. Finally the
real-real piece corresponds to the emission of two additional partons
and its IR singularities correspond to when one, or both partons
become unresolved. Constructing a scheme to regulate these divergences
has been a ongoing task for many
years~\cite{GehrmannDeRidder:2005cm,Catani:2007vq,Czakon:2010td}. Recently
a new regularization scheme, based upon $N$-jettiness
\cite{Stewart:2010tn} has been
proposed~\cite{Boughezal:2015dva,Gaunt:2015pea}. Here the idea is
similar to that used in $q_T$ subtraction~\cite{Catani:2007vq}, and a calculation of the top quark decay at NNLO, 
based on Soft Collinear
Effective Field Theory
(SCET) methods~\cite{Gao:2012ja}.
These
methods introduce a variable which separates the singly unresolved
regions from the doubly unresolved ones. If an all-orders formulation
(i.e. a factorization theorem) is known for the doubly unresolved
region, then an expansion can be performed to a fixed order in the
coupling. The singly unresolved region corresponds to the NLO
calculation of the process with an additional parton, which can be
evaluated using traditional means.  For $q_T$ subtraction, applicable
to production of colour neutral final states, the separation is
obtained via a $q_T$ cut. If $q_T > q_T^{\rm{cut}}$ then the electroweak (EW) system
recoils against a parton, and only single unresolved limits can occur
(i.e. the NLO calculation of the EW final state together with one
additional parton). For $q_T <
q_T^{\rm{cut}}$ the all-orders factorization of Collins, Soper and
Sterman~\cite{Collins:1984kg} can be used. In
refs.~\cite{Boughezal:2015dva,Gaunt:2015pea} $N$-jettiness~\cite{Stewart:2010tn}  was proposed
as the separation-cut i.e. $\tau_N > \tau_N^{\rm{cut}}$ defines a NLO
calculation. When $\tau_N < \tau_N^{\rm{cut}}$ SCET~\cite{Bauer:2000ew,Bauer:2000yr,Bauer:2001yt,Bauer:2001ct,Bauer:2002nz,Stewart:2009yx}
provides a factorization theorem~\cite{Stewart:2010tn,Stewart:2009yx} that can be used to compute the cross section. An advantage of this
method is that it can be applied to coloured final states with jets.  The
recent advances in (a variety of) NNLO regularization schemes has
led to a veritable explosion in the number of phenomenological
predictions at NNLO for $2\rightarrow 2$
scattering~\cite{Currie:2013dwa,Chen:2014gva,Brucherseifer:2014ama,Boughezal:2015dva,Boughezal:2015aha,Boughezal:2015dra,Czakon:2015owf,Ridder:2015dxa,Boughezal:2015ded,Cacciari:2015jma}

% In this paper 

The aim of this paper is twofold. Our chief goal is to provide the first
NNLO calculation including both the DY and $y_t$ contributions, with
full flexibility in the boson decays for both $WH$ and $ZH$
processes. Second, we will apply the recently-developed SCET formalism
to a detailed phenomenological study, including the process
$VH\rightarrow VWW \rightarrow $leptons. Such decays have not
previously been included in NNLO codes, but are studied
experimentally. Given the large and intricate final state phase space
(22 dimensions for the double-real part) this is a particularly good
example to test the feasibility of the SCET regularization to provide
NNLO predictions for complicated phenomenological applications. Our results are
implemented in MCFM~\cite{Campbell:1999ah,Campbell:2011bn,Campbell:2015qma},
and are available in MCFM 8.0~\cite{Boughezal:2016wmq}.

This  paper proceeds as follows. In section~\ref{sec:calc} we present an
overview of the component pieces needed to complete the calculation of $VH$ at
NNLO. Phenomenological results for the LHC Run II are then presented in
section~\ref{sec:pheno}. We draw our conclusions in section~\ref{sec:conc}. We
present a detailed discussion of the helicity amplitudes
needed in the computation of the NNLO correction in Appendices~\ref{app:DYamp}
and~\ref{app:ytamp}.

\section{Calculation} 
\label{sec:calc}

In this section we describe the details of our NNLO calculation for
$VH$ production and its implementation into a fully flexible Monte Carlo
code. The aim of this section is to provide an overview of the
calculation, and its subsequent implementation in MCFM. Technical details regarding the calculation of the 
amplitudes are presented in Appendices~\ref{app:DYamp}-\ref{app:ytamp}.  At NNLO the
production cross-section $d\sigma^{(2)}_{pp\rightarrow \ell_1\ell_2 H}$ 
can be written as the sum of two terms,
\begin{eqnarray}
d\sigma^{(2)}_{pp\rightarrow \ell_1\ell_2 H}= d\sigma^{(2),DY}_{pp\rightarrow \ell_1\ell_2 H}+d\sigma^{(2),y_t}_{pp\rightarrow \ell_1\ell_2 H}
\label{eq:xsdecomp}
\end{eqnarray}
Here the first term represents the contributions which have the same structure
as single vector boson production, the second term represents a new type of
contribution that occurs first at $\mathcal{O}(\alpha_s^2)$. These pieces
arise from terms in which the Higgs boson couples directly to a heavy quark
(predominantly a top-quark).  In the following sections we first describe these two contributions in more detail,
and then discuss our handling of the decays of the Higgs boson.

\subsection{Drell-Yan type contributions} 

\begin{figure}
\begin{center} 
\includegraphics[width=14cm]{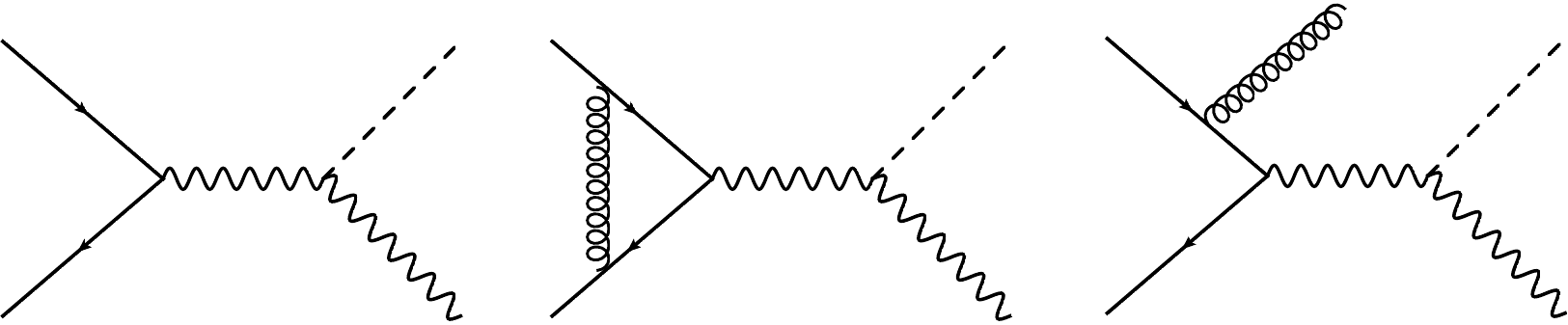} 
\caption{Drell-Yan like production modes for the associated production of a Higgs boson. Shown are representative Feynman diagrams needed 
to compute the $\mathcal{O}(\alpha_S^0)$ (left) and $\mathcal{O}(\alpha_S)$ (center and right) parts of the production cross-section.}
\label{fig:DYNLO}
\end{center} 
\end{figure}

At LO and NLO the production cross-section
$d\sigma^{(i)}_{pp\rightarrow \ell_1\ell_2 H}$ (where $i=0,1)$
has the same structure as the calculation of single vector
boson production. At LO only $q\overline{q}$ initial states contribute,
while the NLO
corrections consist of virtual (one-loop) corrections to this process, and
real-radiation in which the underlying matrix elements contain a
$q\overline{q}$ pair and a gluon. Representative Feynman diagrams for these pieces are illustrated in
Fig.~\ref{fig:DYNLO} where, for simplicity, we have suppressed the decays
of the vector and Higgs bosons.  At NLO IR singularities are isolated using dimensional regularization
and handled using Catani-Seymour dipole subtraction~\cite{Catani:1996vz}.

\begin{figure}
\begin{center} 
\includegraphics[width=14cm]{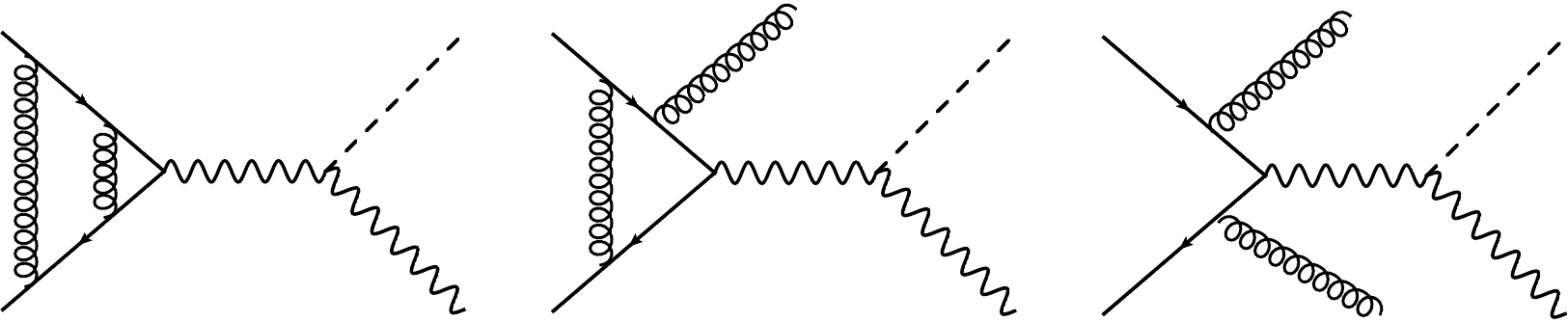} 
\caption{Drell-Yan like production modes for the associated production of a Higgs boson. Shown are representative Feynman diagrams needed 
to compute the $\mathcal{O}(\alpha_S^2)$ corrections to the process. Examples are shown for each of the 0-, 1-, and 2-parton phase space configurations.}
\label{fig:DYNNLO}
\end{center} 
\end{figure}

At NNLO the production cross-section receives contributions from the
$VH + 0,1$ and 2 parton phase spaces. Representative Feynman diagrams
for each of these terms are presented in Figure~\ref{fig:DYNNLO}.
Utilizing the similarities with the NNLO calculation of the Drell-Yan
process~\cite{Hamberg:1990np}, cross-sections for inclusive on-shell
$VH$ production were presented at this order in ref.~\cite{Brein:2003wg}.  At
$\mathcal{O}(\alpha_s^2)$ $d\sigma^{(2),DY}_{pp\rightarrow
  \ell_1\ell_2 H}$ contains UV poles which we renormalize in the
$\overline{\rm{MS}}$ scheme. In addition to the UV divergences,
$d\sigma^{(2),DY}_{pp\rightarrow \ell_1\ell_2 H}$ contains
singularities of IR origin. In order to regularize these we use the
recently developed $N$-jettiness slicing
procedure~\cite{Gaunt:2015pea,Boughezal:2015dva,Boughezal:2015aha,Boughezal:2015ded}. This
procedure uses the $N$-jettiness variable ($\tau_N$)
to divide the NNLO calculation into two pieces based on the value of
$\tau_N$. Below the $\tau$-cutoff parameter the technology of
SCET~\cite{Stewart:2010tn,Stewart:2009yx,Kelley:2011ng,Monni:2011gb,Gaunt:2014xga}
is used to provide a factorization theorem. Above
the $\tau_N$-cutoff the calculation reduces to a NLO computation of the
($VH+j$) process, and can
be evaluated using traditional techniques. In MCFM the IR regularization of the
NLO $VHj$ processes is obtained via the Catani-Seymour dipole formalism~\cite{Catani:1996vz}.
Since the SCET formalism
below the $\tau_N$-cutoff is approximate and subject to power corrections,
the value of $\tau_N$ should be taken
as small as possible. A check of the implementation is
thus obtained by checking the cancellation of the logarithmic pieces
above and below the cut. For our process, which does not contain any
final state jets in the Born phase space, the $\tau_N$-cutoff
procedure is similar to the $q_T$ subtraction technique~\cite{Catani:2007vq}
used in previous calculations~\cite{Ferrera:2011bk}.  A detailed study
of the $N$-jettiness regularization for colour singlet final states and
their implementation in MCFM is presented in ref.~\cite{Boughezal:2016wmq}, to which we
refer the interested reader for more details.  MCFM also contains implementations of
one-jet production in association with a Higgs~\cite{Boughezal:2015aha},
$W$~\cite{Boughezal:2015dva} or $Z$~\cite{Boughezal:2015ded} boson, and diphoton production~\cite{Campbell:2016yrh}.
We stress that in MCFM we only use $N$-jettiness slicing to calculate the coefficient of the $\mathcal{O}(\as^2)$ term 
in the perturbative expansion. 

In order to implement the DY pieces in MCFM we need the two-loop virtual
amplitude~\cite{Hamberg:1990np} interpreted in terms of the hard function of
SCET~\cite{Idilbi:2006dg,Becher:2006mr}, and the NLO implementation of $VH+j$. The results for the two-loop virtual
amplitude are readily available in the literature~\cite{Hamberg:1990np}.
We have calculated the NLO corrections to the $VH+j$ process and implemented
them in MCFM.  Details of the relevant calculational ingredients are
presented in Appendix~\ref{app:DYamp}. 

\subsection{Top Yukawa contributions}

\begin{figure}
\begin{center} 
\includegraphics[width=9cm]{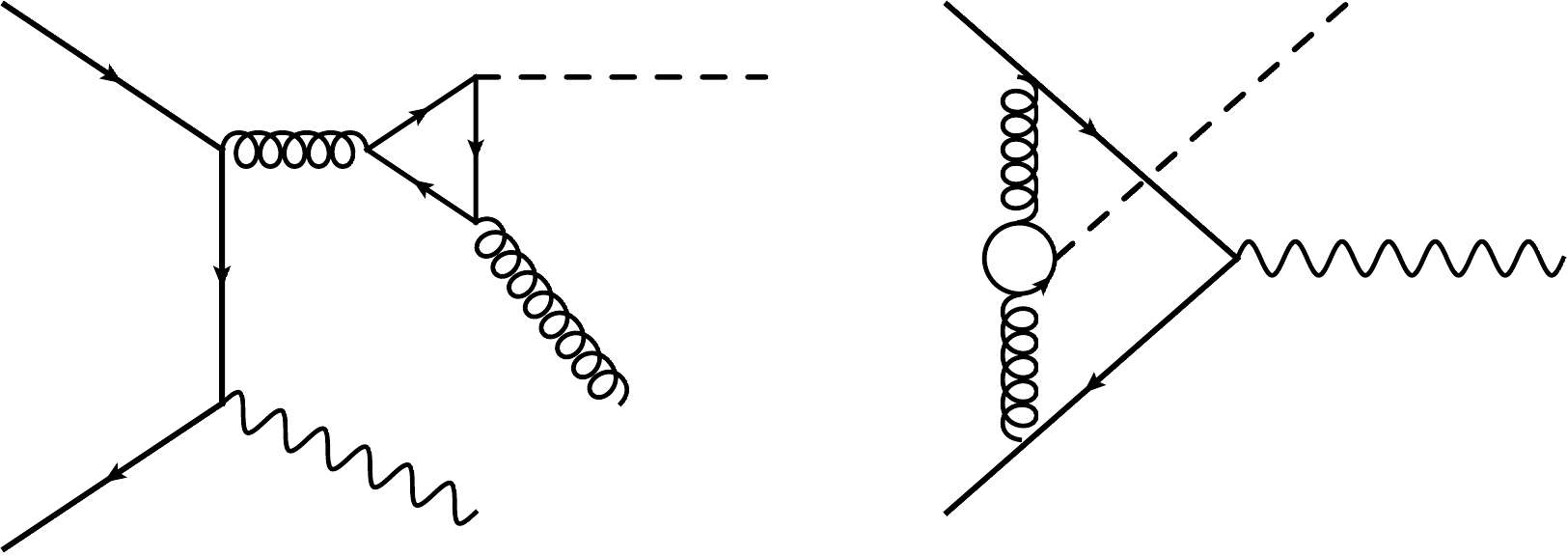} 
\caption{Production modes for the contributions that are proportional to the top Yukawa coupling $y_t$ for the associated production of a Higgs boson.
These topologies occur for either $WH$ or $ZH$ production, and interfere with the LO amplitude. }
\label{fig:WHtop}
\end{center} 
\end{figure}

\begin{figure}
\begin{center} 
\includegraphics[width=12cm]{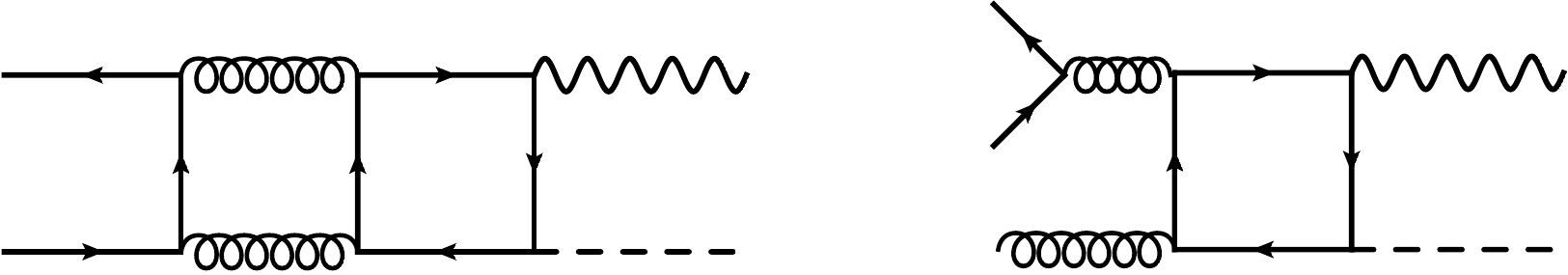} 
\caption{Production modes for the contributions that are proportional to the top Yukawa coupling $y_t$ for the associated production of a Higgs boson.
These types of topology only occur for $ZH$ production.}
\label{fig:ZHtop}
\end{center} 
\end{figure}

\begin{figure}
\begin{center} 
\includegraphics[width=10cm]{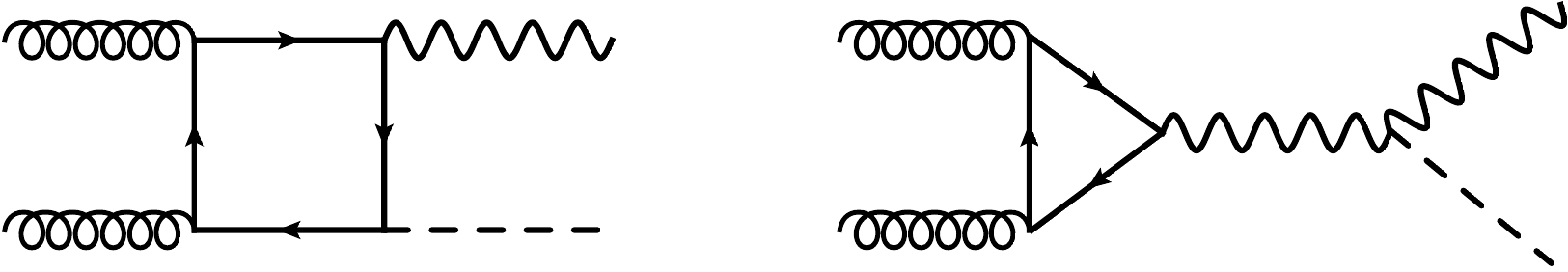} 
\caption{Representative Feynman diagrams for the self-interfering $gg\rightarrow HZ$ contribution. Not all of the diagrams depend on $y_t$ as can be seen from 
the examples on the left ($y_t$) and right (no~$y_t$). 
These topologies only occur for $ZH$ production.}
\label{fig:ggZH}
\end{center} 
\end{figure}

A new type of process opens up at $\mathcal{O}(\as^2)$ and corresponds
to diagrams in which the Higgs boson does not couple directly to the
vector boson, but instead couples to a massive quark. Since the
top quark has by far the largest Yukawa coupling, these contributions
are dominated by the top-quark loops. These $y_t$ diagrams further
sub-divide into two categories. Diagrams of the first kind,
representatives of which are presented in Fig.~\ref{fig:WHtop}, contain
a closed loop of heavy quarks which does not radiate the vector
boson. The second kind, illustrated in Fig.~\ref{fig:ZHtop}, contains
diagrams that include a
closed loop of fermions which radiates both the Higgs and the vector
boson.  Charge conservation mandates that the latter examples are
forbidden if the radiated boson is a $W$. Therefore the first
topologies (Fig.~\ref{fig:WHtop}) occur for both $WH$ and $ZH$ production
and the latter topologies occur only in the $ZH$ case. Both sets of topologies
can have two-loop $q\overline{q}$ topologies, which interfere with the LO
amplitude, and one-loop $q\overline{q} g$ topologies, which interfere
with the $q\overline{q} g VH$ tree amplitude. These pieces have been
computed for on-shell final state particles in ref.~\cite{Brein:2011vx} and we
follow the nomenclature introduced in that paper. We refer to the
two-loop diagrams by the label $V$ and the one-loop diagrams by $R$.
The sub-topologies of these sets are further
distinguished by $I$ (for diagrams that occur for both
$WH$ and $ZH$ production) and $II$ ($ZH$ only).  In ref.~\cite{Brein:2011vx}
these pieces were computed and found to contribute around 1--3$\%$ of
the total NNLO cross-section. Whilst this may appear to be a small
contribution that can safely be neglected, the total NNLO correction
from the DY-type diagrams discussed previously is itself of the same
order. Therefore in order to obtain a reliable prediction at
$\mathcal{O}(\alpha_s^2)$ it is crucial to include both contributions.
Hence a primary aim of this paper is to implement the corrections in this
way in a fully flexible Monte Carlo code.

Finally we observe that for $ZH$ production a gluon-initiated loop arises
that interferes with itself at $\mathcal{O}(\alpha_s^2)$. Example diagrams
are depicted in Fig.~\ref{fig:ggZH}. Due to the enhancement from the gluon
parton distribution function (pdf) these contributions represent a large part of
the NNLO correction, particularly in the boosted regime that is defined
by high vector boson, or Higgs boson, transverse
momentum~\cite{Englert:2013vua}. 
Throughout this paper we include the $gg\rightarrow ZH$ 
diagrams in the $y_t$ contribution. This is a slight abuse of nomenclature
since, as can be seen in Fig.~\ref{fig:ggZH}, there is a triangle diagram corresponding to
$gg \rightarrow Z^* \rightarrow ZH$, where the virtual $Z$ boson
radiates a Higgs boson. The $gg\rightarrow HZ$ contribution
was included in the on-shell prediction of~\cite{Brein:2003wg} and in the more
differential calculation of ref.~\cite{Ferrera:2014lca}. NLO corrections have
been considered in the heavy top limit in ref.~\cite{Altenkamp:2012sx} and
further improved through soft gluon resummation in ref.~\cite{Harlander:2014wda}.

The one-loop processes $R_I$, $R_{II}$ and $gg\rightarrow HZ$ can all be
calculated in the full theory in which the top mass is retained. The
calculation of the two-loop $V_I$ and $V_{II}$ contributions in the
full theory is much more complicated and at present, the master
integrals are not fully known. Therefore in our calculation and that
of ref.~\cite{Brein:2011vx} an asymptotic expansion in $m_t$ is
performed. Since both $V$ and $R$ pieces are separately finite, there
is some freedom in how the top quark is treated in each part of the
calculation.  Our strategy is to include the full top mass effects
where possible and to perform an asymptotic expansion only when
needed.  Finally we note that $R_{II}$ was found to have a very small
effect on the total prediction in~ref.~\cite{Brein:2011vx} so we do
not include it in our calculation of $ZH$.  Technical details
regarding our implementation of these pieces in MCFM are presented in
Appendix~\ref{app:ytamp}.

\subsection{Decays of unstable bosons} 

The aim of this paper is to present a fully flexible Monte Carlo code
for the associated production of a Higgs and vector boson. A crucial
element of this flexibility is to ensure that the relevant decays of
the Higgs and vector bosons are included.
%Phenomenologically speaking the most relevant 
%decays of the vector bosons are those to leptons. 
Decays of vector bosons to leptons represent the cleanest experimental
signature of these processes, so in this paper we focus on the decays
$W\rightarrow \ell\nu$ and $Z\rightarrow \ell^+\ell^-$.  For the Higgs
boson the $b\overline{b}$ decay is the most useful, primarily due to
its high yield rather than its experimental
cleanliness~\cite{Aad:2014xzb,Chatrchyan:2013zna}. However,
$H\rightarrow WW^*$ decays also provide a viable experimental
signature with the current data
set~\cite{ATLAS-CONF-2015-005,CMS:2013xda}. The high luminosity runs
of the LHC may also be able to study rarer channels, such as
$VH\rightarrow \ell\ell\gamma\gamma$. These channels are much cleaner,
since the dominant irreducible backgrounds from $V\gamma\gamma$ are
much smaller, but the small $H\rightarrow \gamma\gamma$ branching
ratio makes detailed studies of this process impractical at present.

The decays discussed above are easily incorporated in Monte Carlo
codes. However, when considering higher-order QCD corrections, the
decay $H\rightarrow b\overline{b}$ requires further discussion since
radiation can occur in both production and decay stages. At NLO
($\mathcal{O}(\alpha_s)$) the conservation of colour ensures a complete
factorization between production and decay processes. The situation is
more complicated at NNLO ($\mathcal{O}(\alpha_s^2)$) since here
contributions exist which connect the initial- and final-state
partons.  It has been shown~\cite{Fadin:1993kt,Fadin:1993dz} that
these these (non-factorizable) pieces contribute to the total rate at
order $\Gamma_H/m_{b\bar{b}}$ where $\Gamma_H$ is the Higgs boson
width.  The situation for differential distributions is less precise,
but it is plausible that for distributions where $b$ and $\bar{b}$ are
not distinguished the non-factorizable contributions should be
similarly small.  In our calculations we therefore neglect such
effects and follow a factorized approach in which the Higgs decay is
included to NLO accuracy.  We follow the procedure outlined in
refs.~\cite{Ferrera:2013yga,Ferrera:2014lca} and define,
\begin{eqnarray}
d\sigma^{\mathrm{NNLO}} \equiv
d\sigma^{\mathrm{NNLO(prod)+NLO(dec)}}_{pp\rightarrow VH \rightarrow \ell_1\ell_2 b\overline{b}} &=& Br(H\rightarrow b\overline{b}) \nonumber
\times\bigg\{ d\sigma^{(0)}_{pp\rightarrow \ell_1\ell_2 H} \times \frac{d\Gamma^{(0)}_{H\rightarrow{b\overline{b}}}+d\Gamma^{(1)}_{H\rightarrow{b\overline{b}}}}{\Gamma^{(0)}_{H\rightarrow b\overline{b}}+\Gamma^{(1)}_{H\rightarrow b\overline{b}}}
\nonumber\\&&+
\left(d\sigma^{(1)}_{pp\rightarrow \ell_1\ell_2 H}+d\sigma^{(2)}_{pp\rightarrow \ell_1\ell_2 H}\right)\times \frac{d\Gamma^{(0)}_{H\rightarrow{b\overline{b}}}}{\Gamma^{(0)}_{H\rightarrow b\overline{b}}}
\bigg\}
\end{eqnarray}
In the above equation $d\sigma^{(i)}_{pp\rightarrow \ell_1\ell_2 H}$ represents the $\mathcal{O}(\alpha_s^i)$ term in the perturbative expansion for the production of
a Higgs boson and a pair of leptons. $d\Gamma^{i}_{H\rightarrow b\overline{b}}$ represents the differential partial width at $\mathcal{O}(\alpha_s^i)$ for the
$H\rightarrow b\overline{b}$ decay, whilst $\Gamma^{i}_{H\rightarrow b\overline{b}}$ represents the integrated partial width for these decays. 
In order to study the effect of the pure NNLO corrections it is also useful to define,
\begin{equation}
d(\Delta\sigma^{\mathrm{NNLO}}) = Br(H\rightarrow b\overline{b}) \times
d\sigma^{(2)}_{pp\rightarrow \ell_1\ell_2 H} \times \frac{d\Gamma^{(0)}_{H\rightarrow{b\overline{b}}}}{\Gamma^{(0)}_{H\rightarrow b\overline{b}}}
\end{equation}
such that $d\sigma^{\mathrm{NLO}} = d\sigma^{\mathrm{NNLO}} - d(\Delta\sigma^{\mathrm{NNLO}})$ defines the prediction that treats
both radiation in production and decay stages at the NLO level.

Radiative corrections to the decay $H\rightarrow b\overline{b}$ were first
computed over thirty years ago~\cite{Braaten:1980yq}. It was shown that there
are large differences between the partial width in a ``massless'' theory, in
which the $b$-quark mass is kept in the Yukawa coupling but dropped in the
matrix element and phase space, and the full theory in which a non-zero
bottom quark mass is retained throughout.
These large differences are the result of logarithms of the form
$\log{(m_b^2/m_H^2)}$ that can be absorbed into a re-definition of
$m_b$ in the Yukawa coupling. As a result, if the running bottom quark mass
is used then the massless and massive predictions are very similar. In our MCFM implementation we
keep the mass of the $b$-quark in full, and do not run the $b$-quark mass in the
LO partial width. In order to ensure that the parts of the cross-section that
are only exposed to a LO partial width are not susceptible  to the running mass
corrections, we divide out the partial width and normalize to the branching
ratio, $BR(H\rightarrow b\overline{b})$.  In this way we can also take advantage
of advanced theoretical predictions for this branching ratio, which
is now known to $\mathcal{O}(\alpha_s^4)$~\cite{Baikov:2005rw}.  In our MCFM implementation 
we use the value obtained from the HDECAY code~\cite{Djouadi:1997yw}. We note that, although we do not
currently include effects beyond NLO in the decay, differential calculations for these
quantities have been presented in the massless theory~\cite{Anastasiou:2011qx,DelDuca:2015zqa}.

\section{LHC Phenomenology }
\label{sec:pheno}

In this section we study the phenomenology of the $VH$ processes at
NNLO for the LHC Run II. For the larger rate $H\rightarrow b\overline{b}$ decay we present results 
which can be compared to data collected with the current operating energy of $\sqrt{s}=13$ TeV. For the rarer 
$H\rightarrow WW^*$ and $H\rightarrow \gamma\gamma$ processes we instead focus on predictions which may be
compared with a larger data set obtained in a future $\sqrt{s}=14$ TeV run.  

Our predictions are obtained using the default MCFM EW scheme, which
corresponds to the following parameter choices: $m_W=80.398$ GeV, $m_Z=91.1876$ GeV,
$\Gamma_W=2.1054$ GeV, $\Gamma_Z =2.4952$ GeV, $G_F=1.16639 \times
10^{-5}$ GeV$^{-2}$ and $m_t=172$ GeV.   These are sufficient to determine
the remaining EW parameters.  We use
$m_b=4.75$~GeV and set the CKM matrix elements $V_{ud} = 0.975$ and
$V_{cs} = 0.222$.  Jets are clustered using the anti-$k_T$ jet algorithm
with distance parameter $R=0.4$. Unless otherwise stated, we use the CT14 pdf
sets~\cite{Dulat:2015mca} matched to the appropriate order in
perturbation theory. Our default renormalization and factorization
scale choice is $\mu_R=\mu_F=\mu_0$ with $\mu_0=m_V+m_H$.
At NNLO the dependence on the unphysical renormalization and factorization scales is rather mild, especially  for
$q\overline{q}$ initiated processes such as those under consideration here. As a result any prescription for estimating the
scale uncertainty, for instance by varying both scales in the same direction or varying them in opposite directions, yields
similar results.  However since the $H\rightarrow b\overline{b}$ decay is computed at NLO, a larger scale dependence for this
decay is observed, with the largest deviations at NNLO (for $WH$) arising from the  case where the scales are varied in opposite
directions.  We will therefore present results obtained with
$ \mu_R = k \mu_0$ and $\mu_F = \mu_0 / k$, with $k = 1/2$ and $k = 2$.
%v2 CW

In general our results will show that, once the scale uncertainties discussed above are taken into account, the results for NNLO cross sections
still do not overlap those for NLO.
This is consistent with other NNLO studies of processes that only receive contributions through $q\overline{q}$ initial states at LO,
since the gluon parton distribution is dominant at the relevant partonic energy fractions of the LHC. At LO there is only a very mild scale dependence which 
is completely induced by the factorization scale in the parton distribution functions. At NLO the cross section becomes sensitive to 
the renormalization scale, but typically there is an accidental cancellation between the renormalization and factorization scales 
resulting in a weak scale dependence even at NLO~\cite{Campbell:2011bn}. Therefore interpreting the scale variation as indicative 
of the total theoretical error is unwise at NLO. It is difficult, without knowledge of the N$^3$LO cross section, to predict whether the scale 
variation at NNLO will incorporate higher order predictions. However there is reason to believe this may be the case.  Firstly the process has access to 
all initial state configurations, so there will be no new partonic channels at N$^3$LO. Secondly the recently-reported calculation of the Higgs cross section 
at N$^3$LO~\cite{Anastasiou:2015ema,Anastasiou:2016cez}, is within the scale variation of the NNLO cross section for the first time in the perturbative
expansion. Therefore we are reasonably confident that our scale variation can be interpreted as an indicator of theoretical uncertainty.
As we look at more exclusive quantities, such as cross sections differential in the number of associated jets, this argument begins to break down
and scale variation should not be taken as a rigorous estimate of the theoretical uncertainty. %CW add refs?

A detailed study of color-singlet production (including $VH$
processes) using $N$-jettiness slicing in MCFM 8 is presented in
ref.~\cite{Boughezal:2016wmq}. We refer the interested reader to the detailed discussion of the methodology in that paper and instead briefly summarize the 
checks here. Starting from the $0$-jettiness of a parton $k$ with momentum $p_k$,
\beq
\tau_0(p_k)=\min_{i=a,b}\left\{\frac{2\,q_i\cdot p_k}{E_i}\right\}\ ,
\eeq
where $E_a$, $E_b$ are the energies of the beams~\cite{Stewart:2010tn},
we define the $0$-jettiness as the sum
over all the $M$ final state parton jettiness values,
\beq
\tau_0=\sum_{k=1}^M\tau_0(p_k)=\sum_{k=1}^M
\min_{i=a,b}\left\{\frac{2\,q_i\cdot p_k}{E_i}\right\}\ . 
\eeq
$M$ takes the values $0$, $1$ and $2$, depending on the particular phase space component of the NNLO calculation. 
We then define the above cut region as $\tau_0 > \tau^{\mbox{\tiny{cut}}}$ and the below cut region as $\tau_0 < \tau^{\mbox{\tiny{cut}}}$. 
The terms above and below the cut combine to leave a residual dependence 
on $\tau^{\mbox{\tiny{cut}}}$ that takes the form, 
\begin{equation}
\Delta \sigma^{NNLO}_{\mbox{\tiny jettiness}} (\tauzerocut) =
  \Delta\sigma^{NNLO}
  + c_3\left(\frac{\tauzerocut}{Q}\right)\log^3\left(\frac{\tauzerocut}{Q}\right)
  + c_2\left(\frac{\tauzerocut}{Q}\right)\log^2\left(\frac{\tauzerocut}{Q}\right)
   \;,
\label{eq:nnloform}
\end{equation}
in the limit that $\tau^{\mbox{\tiny{cut}}}/Q\rightarrow 0$.
Here $Q$ defines a hard scale in the LO process (for us $m_V+m_H$)~\cite{Boughezal:2016wmq} and $c_2$ and $c_3$ are coefficients
that can be fitted numerically if desired.  As a test of our implementation, we have validated our calculation in the absence of any
cuts on the final state particles by comparison with the public code {\tt{vh@nnlo}}~\cite{Brein:2003wg,Brein:2012ne}.
In the limit $\tau^{\mbox{\tiny{cut}}}\rightarrow 0$ the two are in perfect agreement. We also note that we 
have checked the calculation of the $y_t$ contributions for on-shell bosons with {\tt{vh@nnlo}}, also finding perfect agreement. 
As a detailed discussion of the {\tt{vh@nnlo}} checks are provided in 
ref.~\cite{Boughezal:2016wmq} we instead focus on similar fits for the phenomenologically relevant processes, in which 
bosonic decays are included, in the following section. The routines for decaying the Higgs boson in MCFM are well established 
and have been checked against calculations of branching ratios available in the literature.

\subsection{Results: $H\rightarrow b\overline{b}$}
\label{sec:Hbb}

In this section we present our results for LHC phenomenology for the $H\rightarrow b\overline{b}$ decay. 
We will study a variety of phase space selection criteria, with cuts inspired by the ATLAS~\cite{Aad:2014xzb} and CMS~\cite{Chatrchyan:2013zna}
experiments. We define the following set of basic cuts,
\begin{eqnarray} 
{\rm{Jets : }}&& \quad  p_T^j > 25 \; {\rm{GeV}}, \;  |\eta_{j} | < 2.5  \label{eq:basid1}
 \\
{\rm{Leptons :}}&& \quad p_T^{\ell} > 25 \; {\rm{GeV}}, \; |\eta_{\ell} | < 2.5 
 \\
{{WH :}}&& \quad \slashed{E}_T > 20 \;  {\rm{GeV}},  \; m_T^W  < 120 \; {\rm{GeV}} \\
{{ZH :}}&& \quad 80 <  \; m_{\ell\ell}  < 100 \; {\rm{GeV}} \label{eq:basid2}
\end{eqnarray}

\begin{figure}
\begin{center} 
\includegraphics[width=12cm]{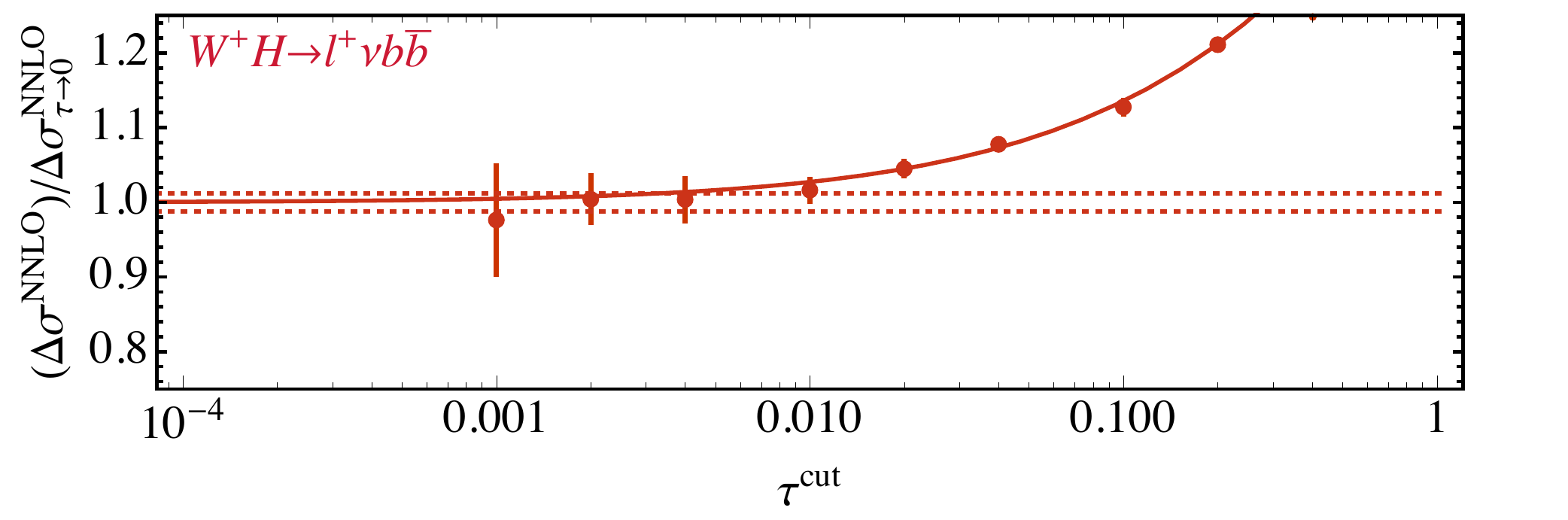} 
\includegraphics[width=12cm]{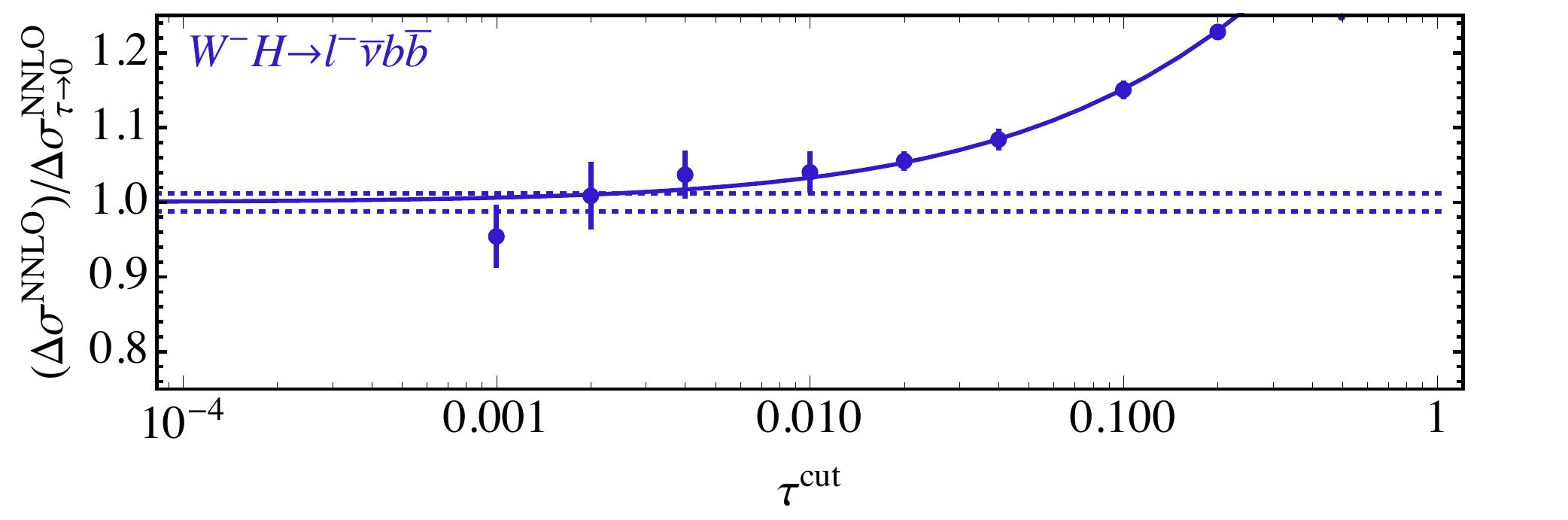} 
\includegraphics[width=12cm]{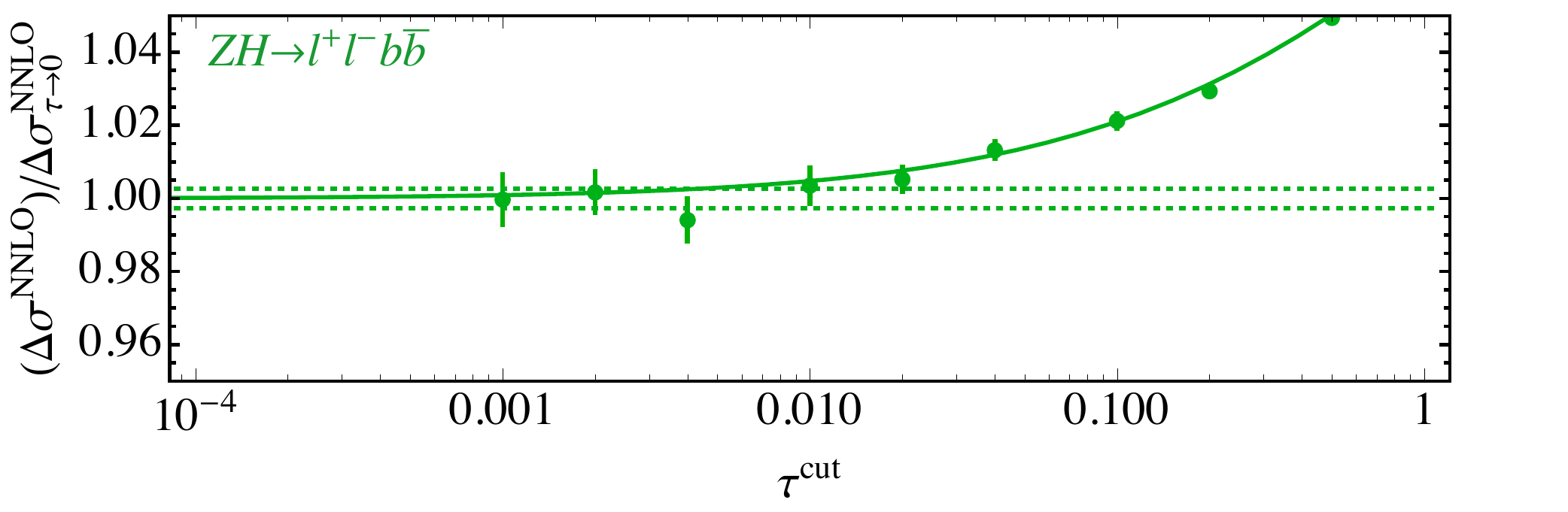} 
\caption{The $\tau$-dependence of the NNLO coefficient for the $VH$ processes,
under the $H \to b{\bar b}$ cuts of Section~\ref{sec:Hbb}.}
\label{fig:tauHbb}
\end{center} 
\end{figure}

Before proceeding we first examine the dependence on the $\tau^{\mbox{\tiny{cut}}}$ parameter 
in the context of the cuts specified in Eqs.~(\ref{eq:basid1})--(\ref{eq:basid2}).
The $\tau^{\mbox{\tiny{cut}}}$-dependence of the NNLO coefficient in the expansion of
the cross-section under these cuts ($\Delta\sigma^{NNLO}$), computed for the $14$~TeV LHC,  is shown in
Figure~\ref{fig:tauHbb}.  The remaining dependence on $\tau^{\mbox{\tiny{cut}}}$ is a result of power corrections,
whose form is given in Eq.~(\ref{eq:nnloform}) in the previous section. The dashed lines indicate the fitting errors
on the asymptotic result for $\tau^{\mbox{\tiny{cut}}} \to 0$.  It is clear that for $WH$ production
$\Delta\sigma^{NNLO}$ is independent of $\tau^{\mbox{\tiny{cut}}}$ at the level of a few
percent for $\tau^{\mbox{\tiny{cut}}} \lesssim 0.01$.  For $ZH$ production the same
value of $\tau^{\mbox{\tiny{cut}}}$ yields an accuracy of about $0.5\%$ in the
coefficient, where, as we will see shortly, the improvement is due to the fact that this process receives
a larger $y_t$ contribution that does not depend on $\tau^{\mbox{\tiny{cut}}}$.
The accuracy of the prediction for the NNLO cross-section
can be assessed by combining this information with the order-by-order
results that are shown in Table~\ref{table:VHbb}.  It is thus clear that
choosing $\tau^{\mbox{\tiny{cut}}}=0.01$ is sufficient for per-mille accuracy in the full
NNLO prediction for all processes.  We shall make this choice henceforth.
\begin{table}
\begin{center}
\begin{tabular}{|c|c|c|c|c|}
\hline
Process & $\sigma_{LO} $ [fb] & $\sigma_{NLO}$ [fb] & $\sigma_{NNLO}$ [fb] & $\Delta\sigma_{NNLO}$ [fb]  \\
\hline\hline 
$W^+H\rightarrow \ell^+\nu b \bar b$     & 19.79 & 20.18 & 20.71 & 0.52 \\
\hline
$W^-H\rightarrow \ell^-\bar\nu b \bar b$ & 14.14 & 14.24 & 14.57 & 0.33 \\
\hline
$ZH\rightarrow \ell^-\ell^+ b \bar b$    & 5.05  & 5.11  & 5.94  & 0.83 \\
\hline 
\end{tabular}
\caption{Cross-sections for $VH$ processes, with leptonic decay of the vector bosons and $H \to b{\bar b}$,
at the 14 TeV LHC. Results are presented for a single family of leptons and correspond to the cuts described
in the text.  Note that, in this table, all cross-sections are computed using the CT14 NNLO pdf set.}
\label{table:VHbb}
\end{center}
\end{table}

\begin{figure}
\begin{center} 
\includegraphics[width=\textwidth]{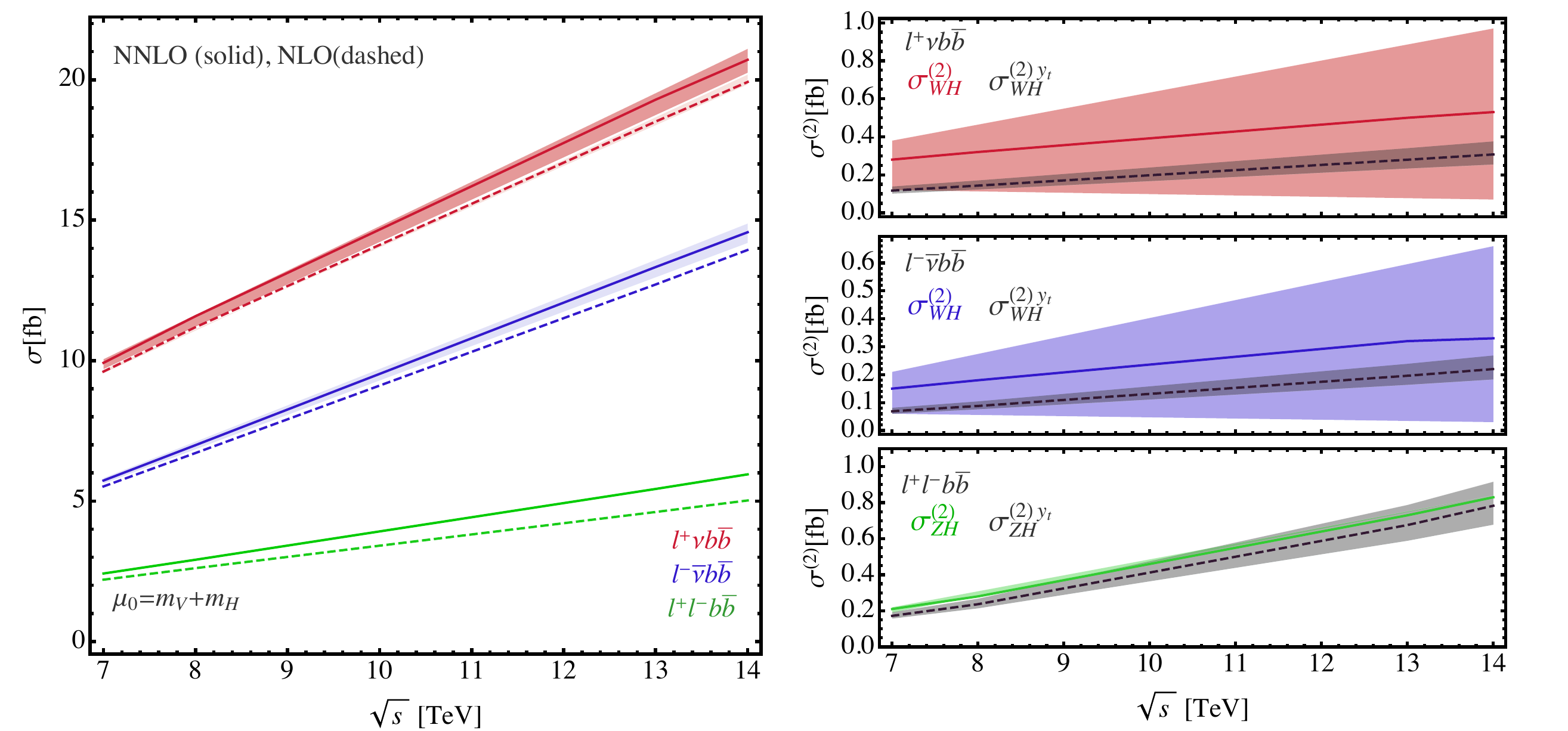} 
\caption{The cross-section in femtobarns as a function of the operating energy $\sqrt{s}$ of the LHC. The basic selection
cuts described in the text are applied. The right-hand plots  show the size of the NNLO coefficient, on each plot the
contribution from the top induced couplings is shown in grey. The shaded bands illustrate the scale-dependence,
computed as described in the text.}
\label{fig:xsbas}
\end{center} 
\end{figure}

In Fig.~\ref{fig:xsbas}, we present the cross-section as a function of
the LHC operating energy given the basic selection cuts described
above. The left-hand plot illustrates the total rate at NLO
(dashed) and NNLO (solid) for $WH$ and $ZH$ production.
We plot the cross-section for $W^+(\rightarrow \ell^+\nu)H$ and
$W^-(\rightarrow \ell^-\overline{\nu})H$ separately.  At the LHC the
production of $W^+H$ is dominant, since the $u\overline{d}$ initial state
configuration has a larger flux than $d\overline{u}$ for $pp$
collisions. The plots on the right hand side show the NNLO
coefficient, $\Delta\sigma^{NNLO}$. The upper and middle panels present
results for $W^+H$ and $W^-H$ production. The NNLO corrections in both
cases are similar. It is interesting to compare the size of the top
induced cross-section $\sigma^{(2),y_t}_{VH}$ to the total NNLO
coefficient. For $WH$ production the top induced pieces, after
cuts, make up around 30-50\% of the total $\mathcal{O}(\alpha_s^2)$
correction.

The $Z$ boson has a much smaller branching ratio to a single family of
leptons compared to the $W$, so the cross-sections presented in
Fig.~\ref{fig:xsbas} for $\ell^+\ell^- b\overline{b}$ are smaller than
the corresponding $W$ induced ones (for instance, compared to the
inclusive results presented in ref.~\cite{Boughezal:2016wmq}). The NNLO
corrections are also much more important for $ZH$ production than for $WH$. This
is due to the large contribution from the $gg\rightarrow ZH$
pieces. The importance of the gluon flux at the LHC 
can help to offset the $\alpha_S$ suppression, resulting in a NNLO correction
whose impact is more comparable to a NLO effect. This is
clearly visible in the lowest plot on the right-hand side of
Fig.~\ref{fig:xsbas}, in which only the NNLO coefficient is
shown. By far the dominant source of the correction arises from
$\sigma^{(2),y_t}_{ZH}$ and not $\sigma^{(2),DY}_{ZH}$.  Of the
$\sigma^{(2),y_t}_{ZH}$ contribution the dominant effect is induced by the $gg$
diagrams, although it is not possible to separate them from the
$V_{i}$ and $R_{i}$ pieces at this order in perturbation theory.

\begin{figure}
\begin{center} 
\includegraphics[width=0.49\textwidth]{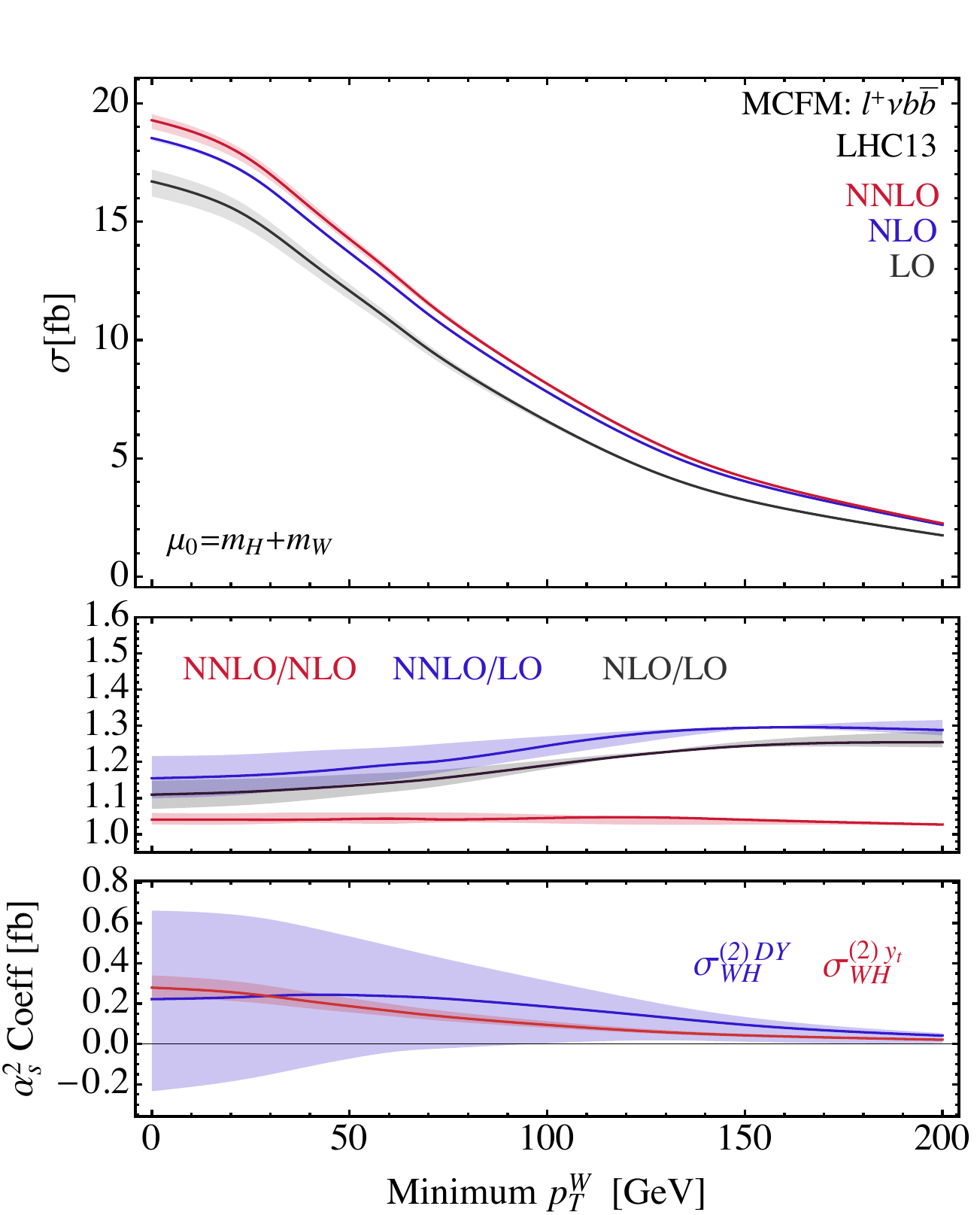} 
\includegraphics[width=0.49\textwidth]{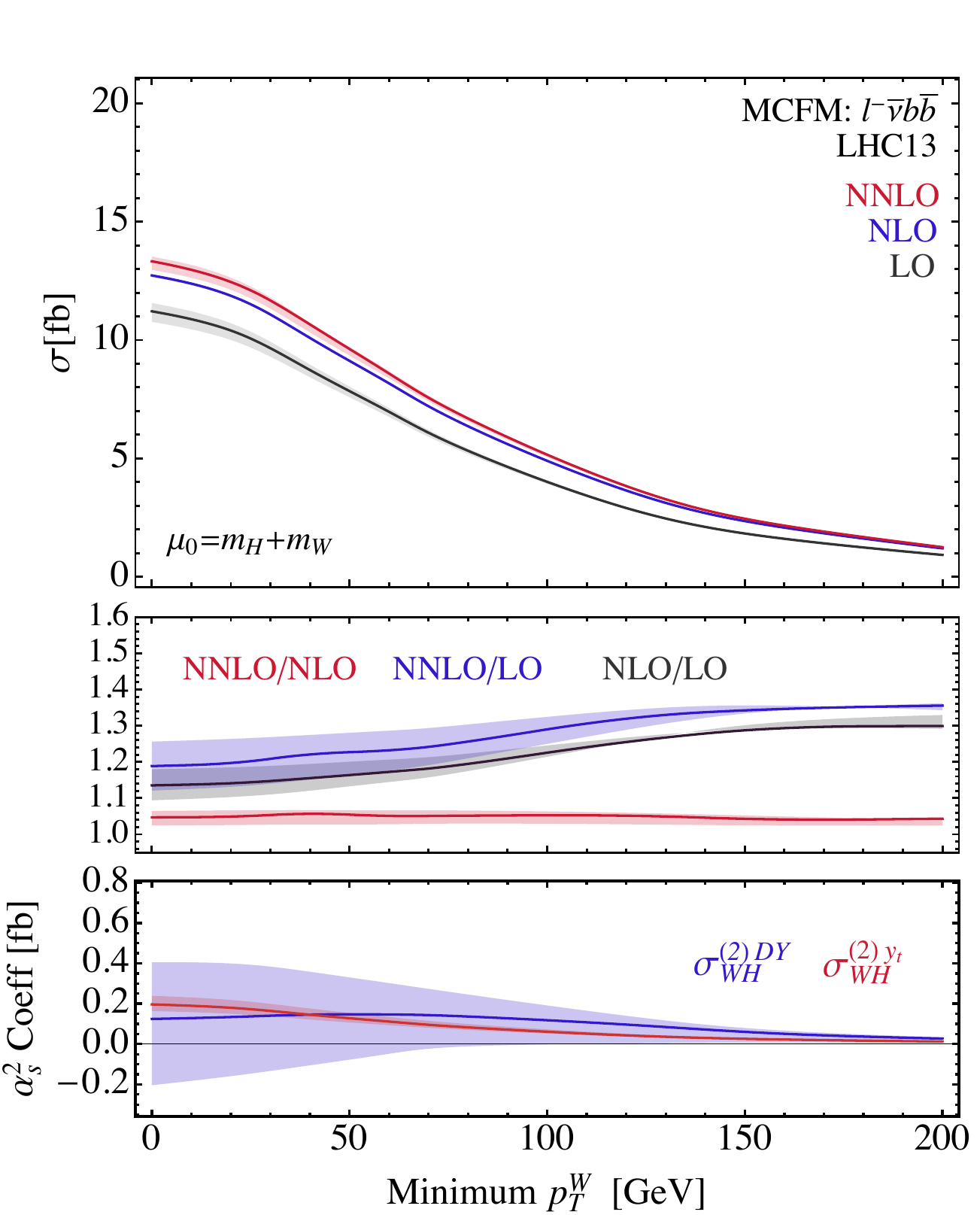} 
\caption{The cross-section in femtobarns as a function of the minimum transverse momentum of the $W^+$ (left) and $W^-$ (right) boson, $p_T^W$. The upper panel presents the total cross-section, 
the middle panel presents the impact of the higher order corrections, the lower plot presents the $DY$ and $y_t$  $\alpha_S^2$ coefficients.}
\label{fig:ptWxs}
\end{center} 
\end{figure}

\begin{figure}
\begin{center} 
%includegraphics[width=0.5\textwidth]{ptV1Z.pdf} 
\includegraphics[width=0.5\textwidth]{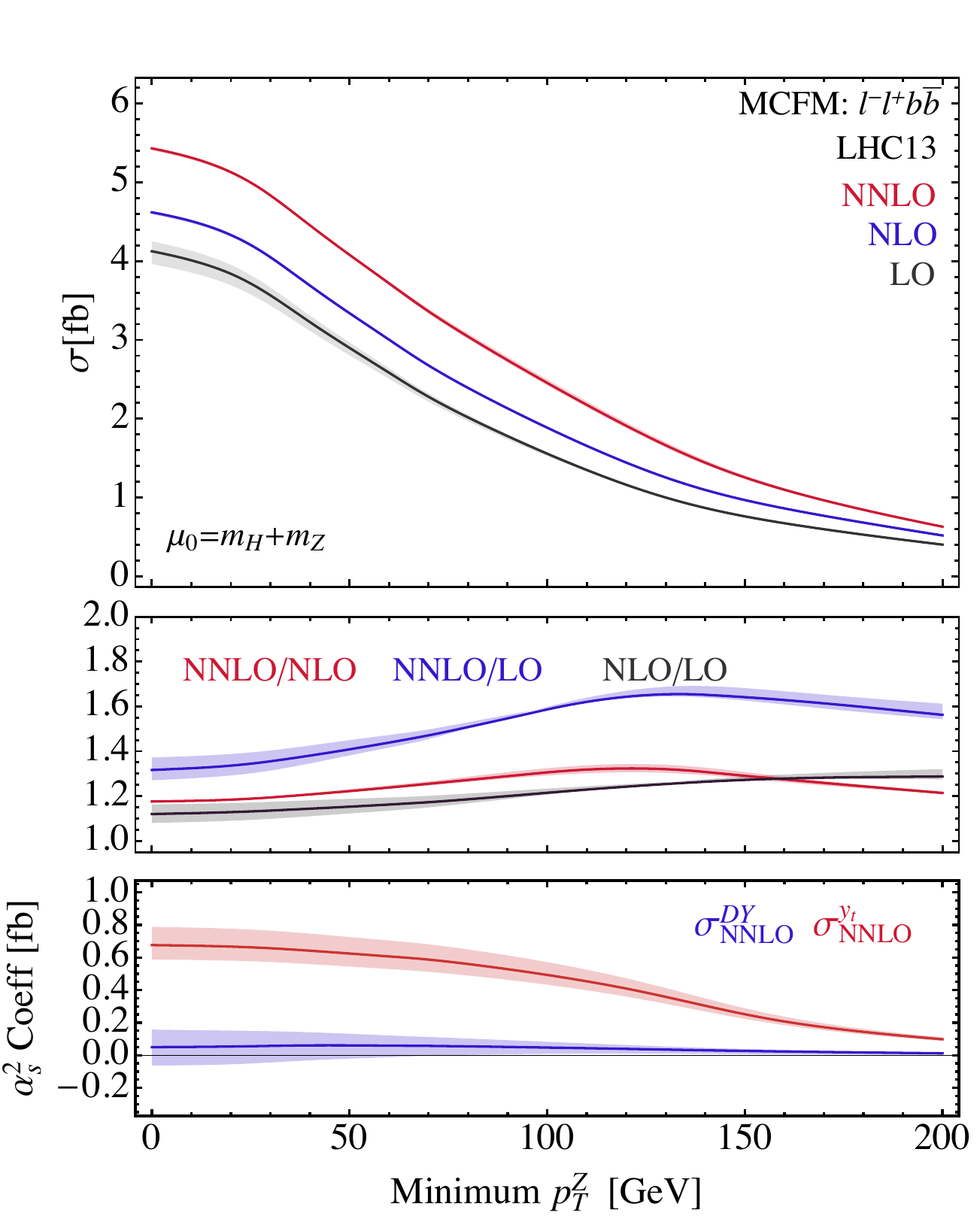} 
\caption{The cross-section in femtobarns as a function of the minimum transverse momentum of the $Z$ boson $p_T^Z$.  The upper panel presents the total cross-section, 
the middle panel presents the impact of the higher order corrections, the lower plot presents the total $\alpha_S^2$ coefficient.}
\label{fig:ptZxs}
\end{center} 
\end{figure}

The cuts described above result in a fairly inclusive selection. In
order to reduce the backgrounds from top, diboson and $V+$ jets processes, cuts
on the transverse momentum of the vector boson are usually employed in
experimental analyses. We therefore investigate the total cross
section as a function of the minimum transverse momentum of the vector
boson in Figs.~\ref{fig:ptWxs} $(WH)$ and~\ref{fig:ptZxs} $(ZH)$.  We
focus on the LHC operating at $\sqrt{s}=13$ TeV. To produce these
results we apply the basic cuts described above. The results of
the previous figures are also manifest in these plots: the initial
impact of higher order corrections for $W^-H$ is slightly larger (at
NLO), but the impact of the NNLO corrections is similar for both
charges in $W^{\pm}H$.  It is also clear that $ZH$ has much larger
NNLO corrections than $WH$. Particularly rich signal bins in the
experimental analysis correspond to $p_T^V > 120$~GeV and $p_T^V > 160$~GeV.
For these choices the signal cross-section is around 30-40\% and $15-20\%$ of the $p_T^V$-inclusive result,
respectively. The impact of NLO is a mild
enhancement in the tail of the $p_T^V$ distribution for all process.  For
$WH$ production the NNLO corrections are reasonably flat in $p_T^V$,
while the NNLO corrections to the $ZH$ process become more pronounced in the high $p_T^V$ region.
This is due almost exclusively to the $y_t$ correction, which hardens the spectrum as can be clearly seen in the middle panel of Fig.~\ref{fig:ptZxs}.

\begin{figure}
\begin{center} 
\includegraphics[width=0.48\textwidth]{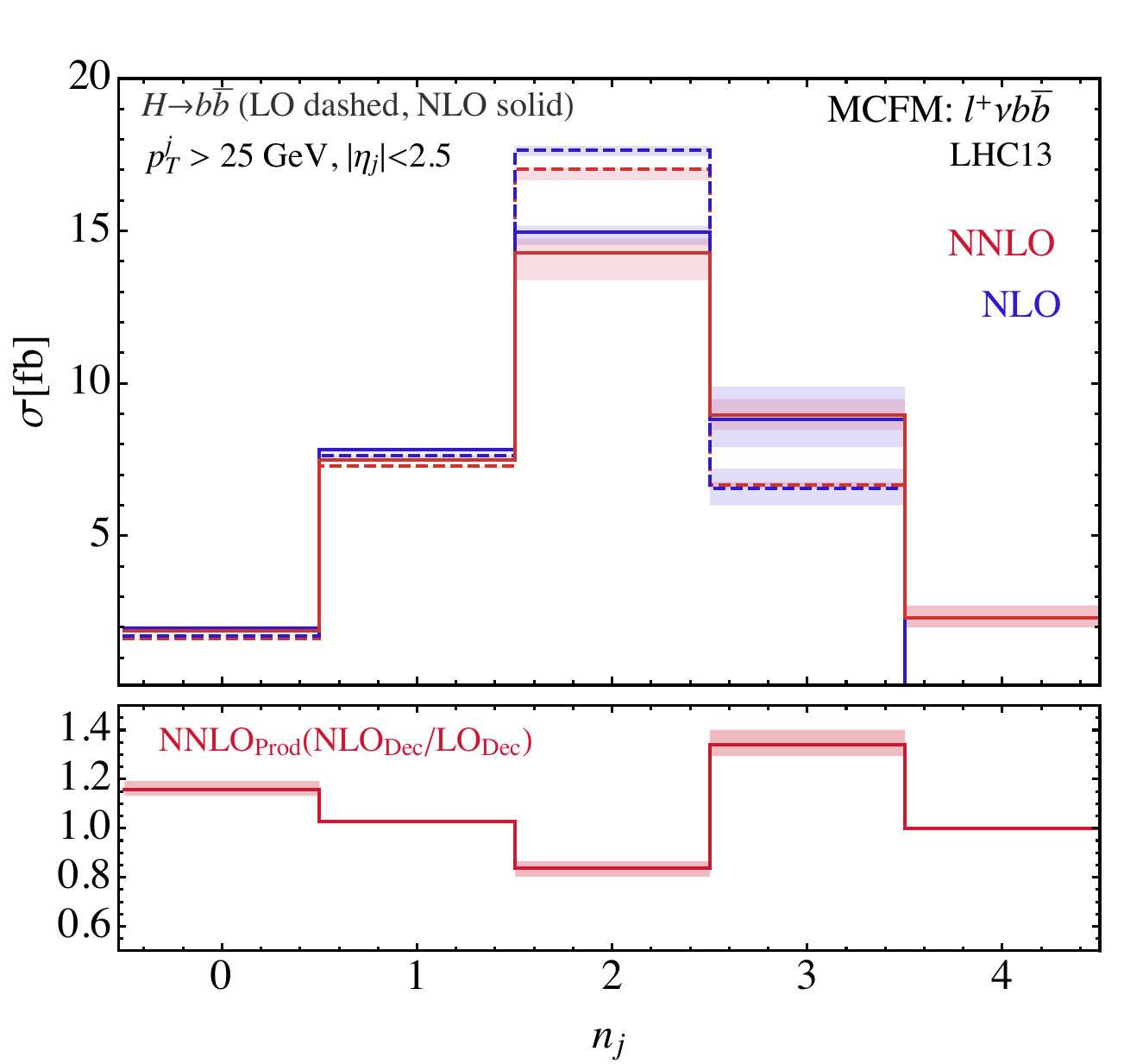} 
\includegraphics[width=0.48\textwidth]{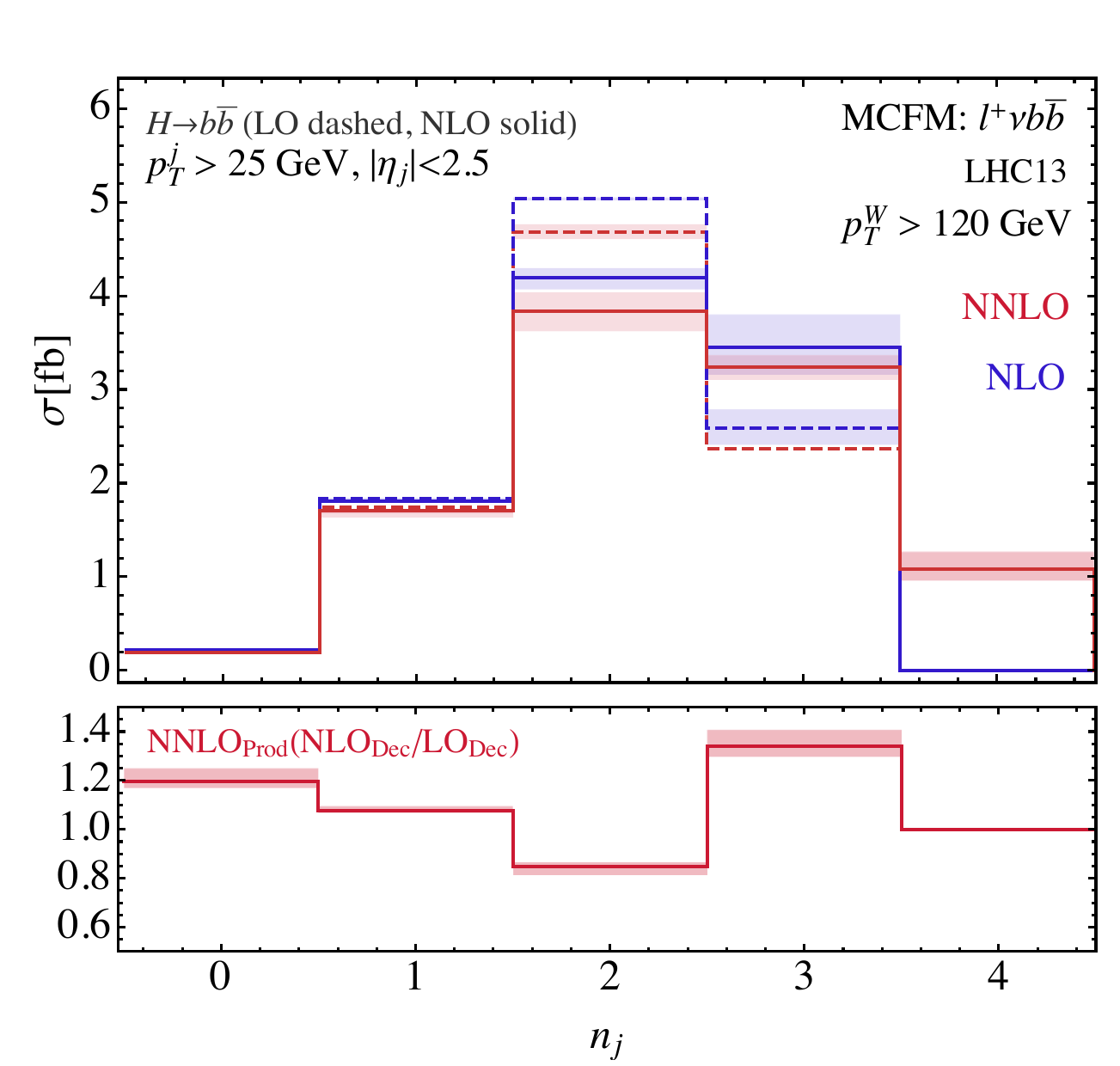} 
\caption{The cross-section in femtobarns as a function of the number of jets (light plus $b$-jets) for $W^+H$ at 13 TeV. The solid lines represent predictions which include 
the $H\rightarrow b\overline{b}$ decay at NLO. }
\label{fig:njetsWp}
\end{center} 
\end{figure}

\begin{figure}
\begin{center} 
\includegraphics[width=0.48\textwidth]{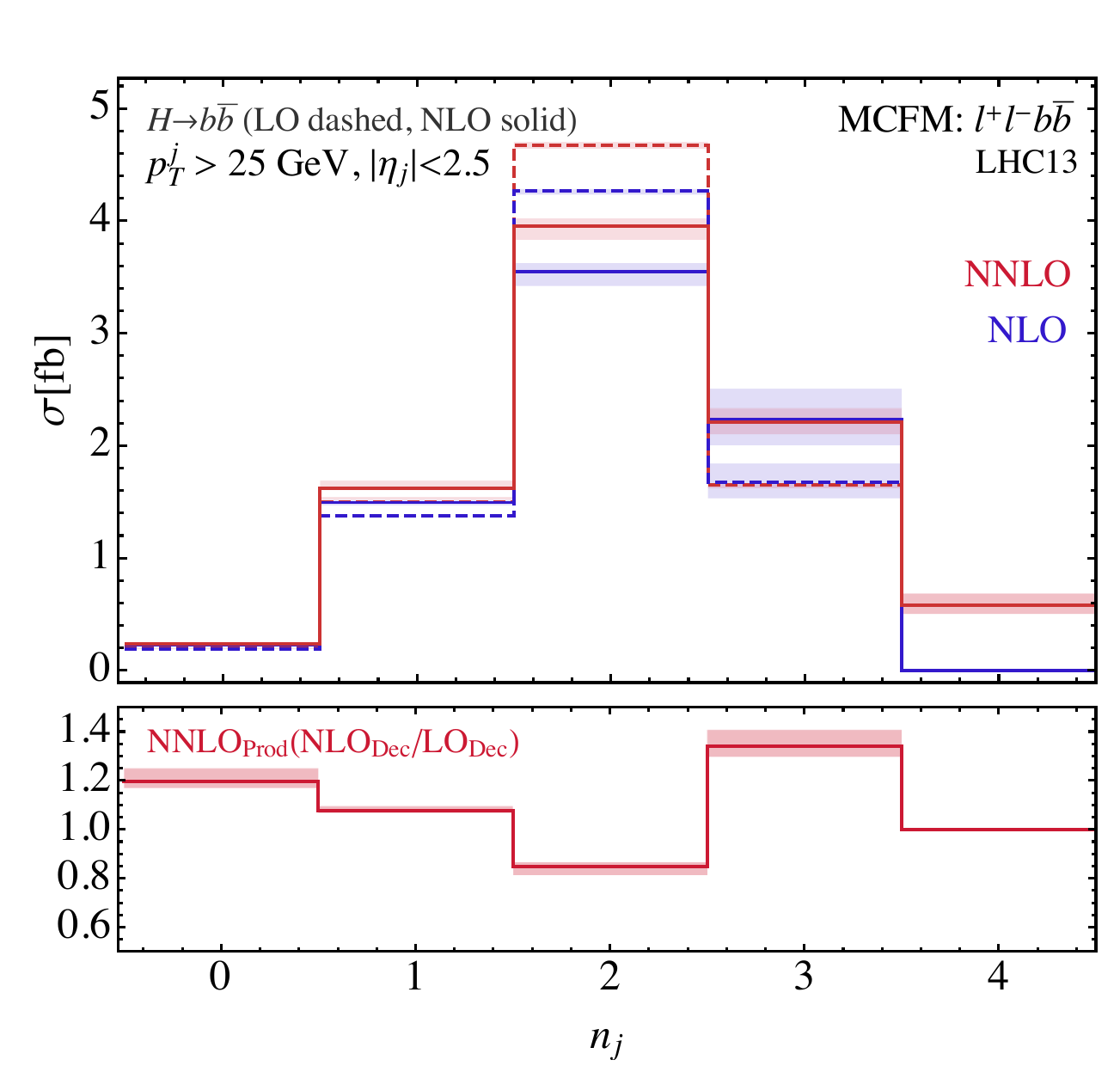} 
\includegraphics[width=0.48\textwidth]{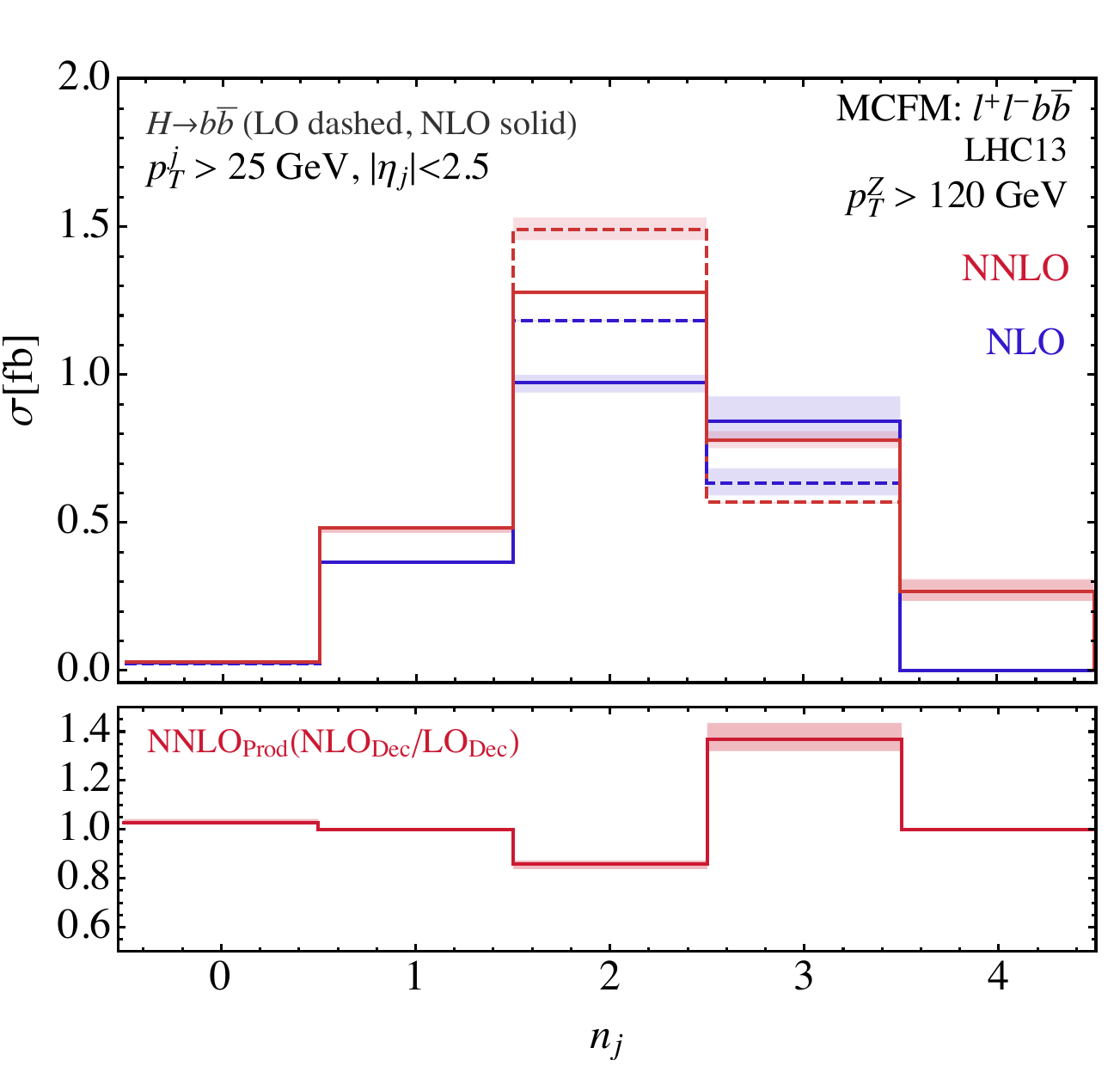} 
\caption{The cross-section in femtobarns as a function of the number of jets (light plus $b$-jets) for $ZH$ at 13 TeV. The solid lines represent predictions which include 
the $H\rightarrow b\overline{b}$ decay at NLO.  }
\label{fig:njetsZ}
\end{center} 
\end{figure}

We now turn our attention to jet-based observables. In
Figure~\ref{fig:njetsWp} we present the cross-section as a function of
the total number of jets (i.e. $b$-jets plus light jets). The plot on
the left-hand side has only the basic lepton cuts applied, while on
the right $p_T^V > 120$~GeV is required in addition to the basic
lepton cuts.  Since the Higgs boson is a resonance decaying to massive
quarks, a well-defined cross-section can be computed without any
requirement on the number of $b$-jets present.  An N$^{n}$LO
prediction can then have between 0 and $(n+2)$ jets in the final
state, with the $(n+2)$-jet bin corresponding to the LO prediction for
$VH+n$ jets. In Fig.~\ref{fig:njetsWp} we present NLO and NNLO
predictions, with NLO (solid) and LO (dashed) $H\rightarrow
b\overline{b}$ decays. As expected the largest scale variation occurs
in the four-jet bin (NNLO) and three-jet bin (NLO) predictions, since
these are LO predictions in this observable. Including the decay has a
significant impact on the jet counting, particularly in the two- and
three-jet bins where it changes the predicted rate by
$\mathcal{O}(20\%)$. The higher $p_T^V$ selection has relatively more
three-jet events than the more inclusive selection, which arises from
the kinematic favorability of balancing a high $p_T$ vector boson with
a jet and the Higgs boson. The Higgs decay at LO has no dependence on
$\alpha_S$ and therefore no scale dependence. When we introduce the
decay at NLO we include $\alpha_S$ for the first time, and acquire a
larger dependence on the choice of scale.  Reducing the scale
dependence further requires consideration of NNLO effects in the decay
stage~\cite{Anastasiou:2011qx,DelDuca:2015zqa}.

\begin{figure}
\begin{center} 
\includegraphics[width=0.48\textwidth]{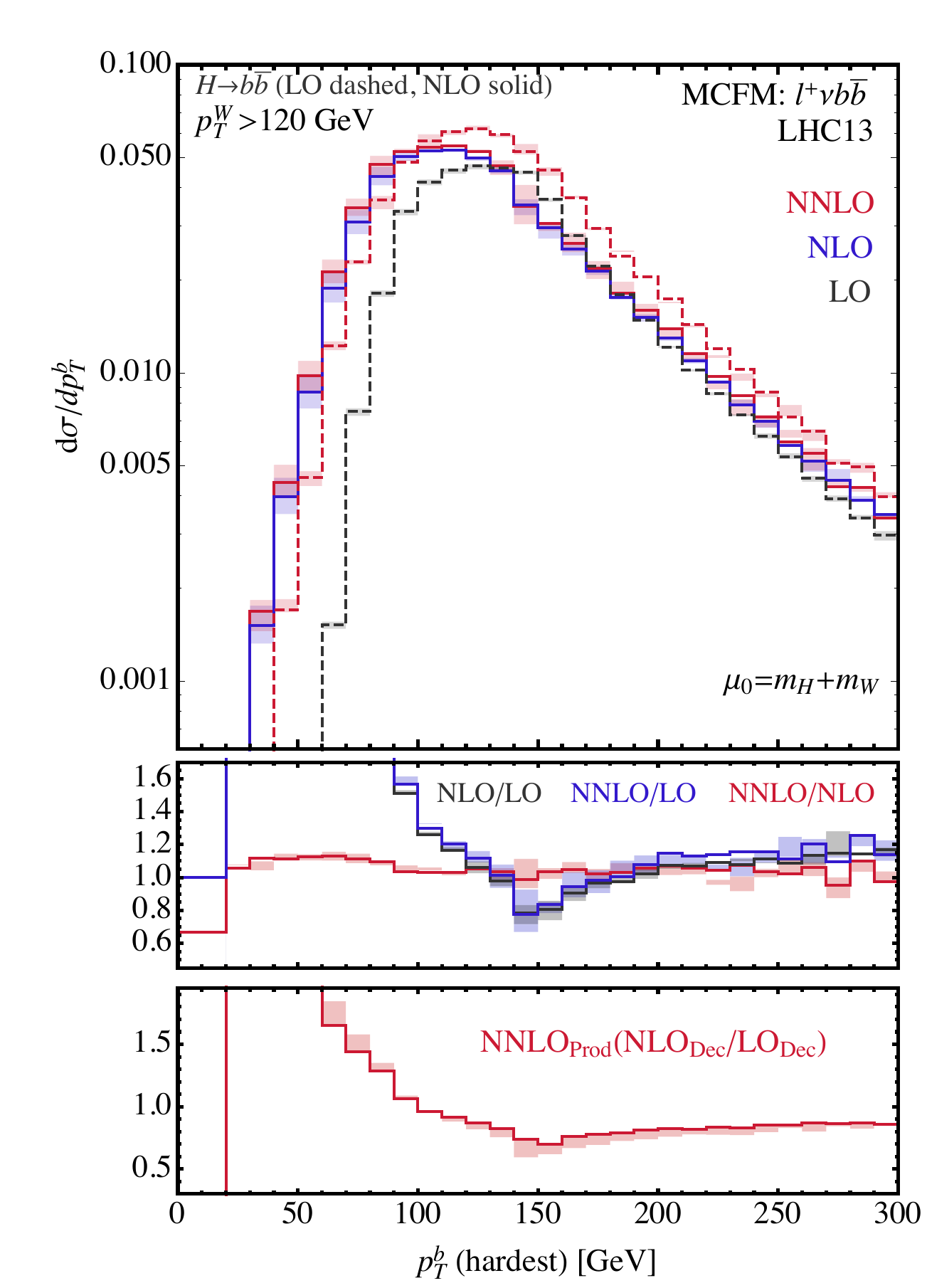} 
\includegraphics[width=0.48\textwidth]{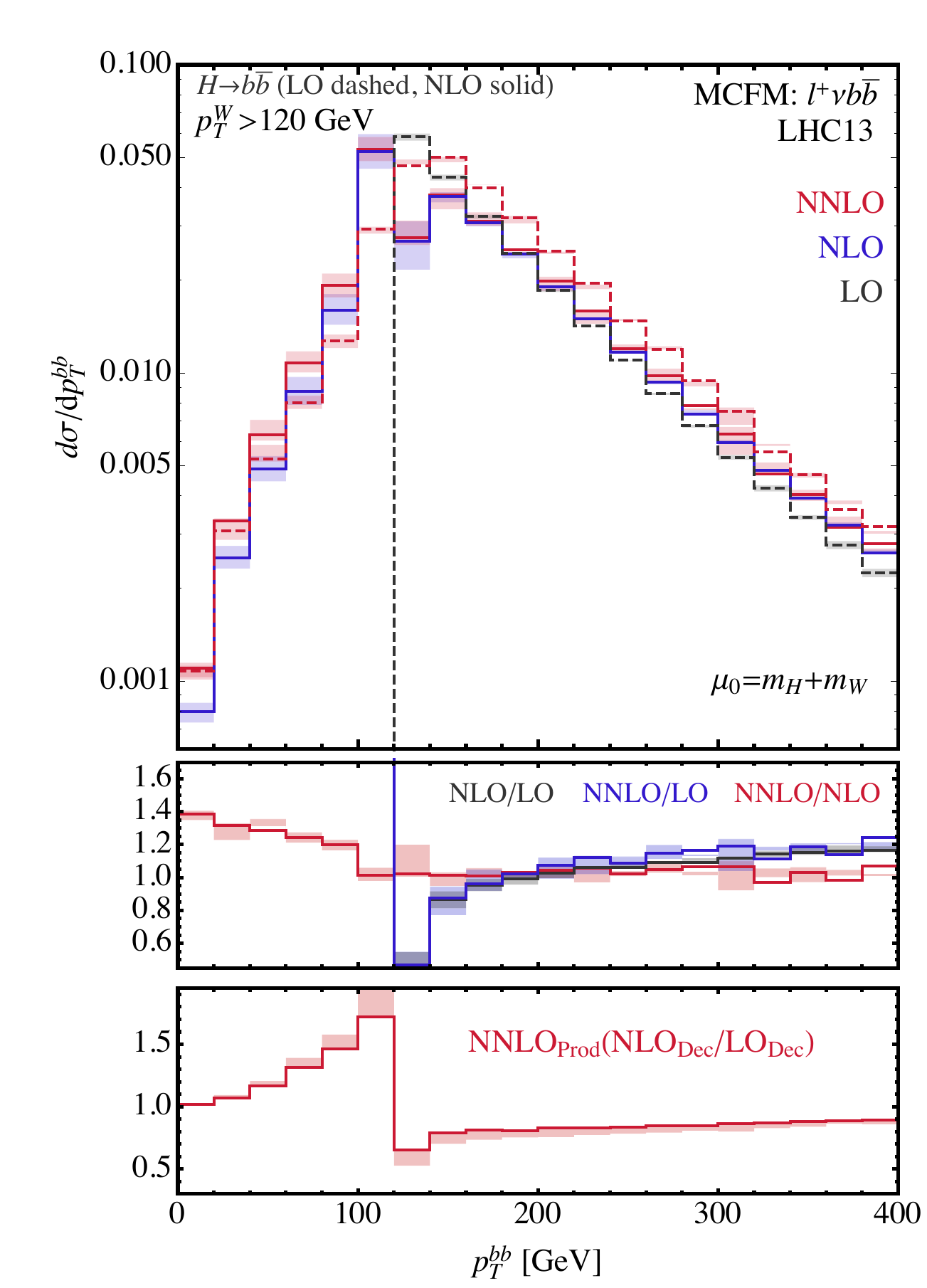} 
\caption{Differential predictions for the transverse momenta of the hardest $b$ (left) and the $b\overline{b}$ system (right)
for $W^+H$ at the LHC. Predictions at NNLO in production with NLO decays are denoted by solid lines, while those with LO decays
are illustrated with dashed curves. }
\label{fig:Wbplots}
\end{center} 
\end{figure}

\begin{figure}
\begin{center} 
\includegraphics[width=0.48\textwidth]{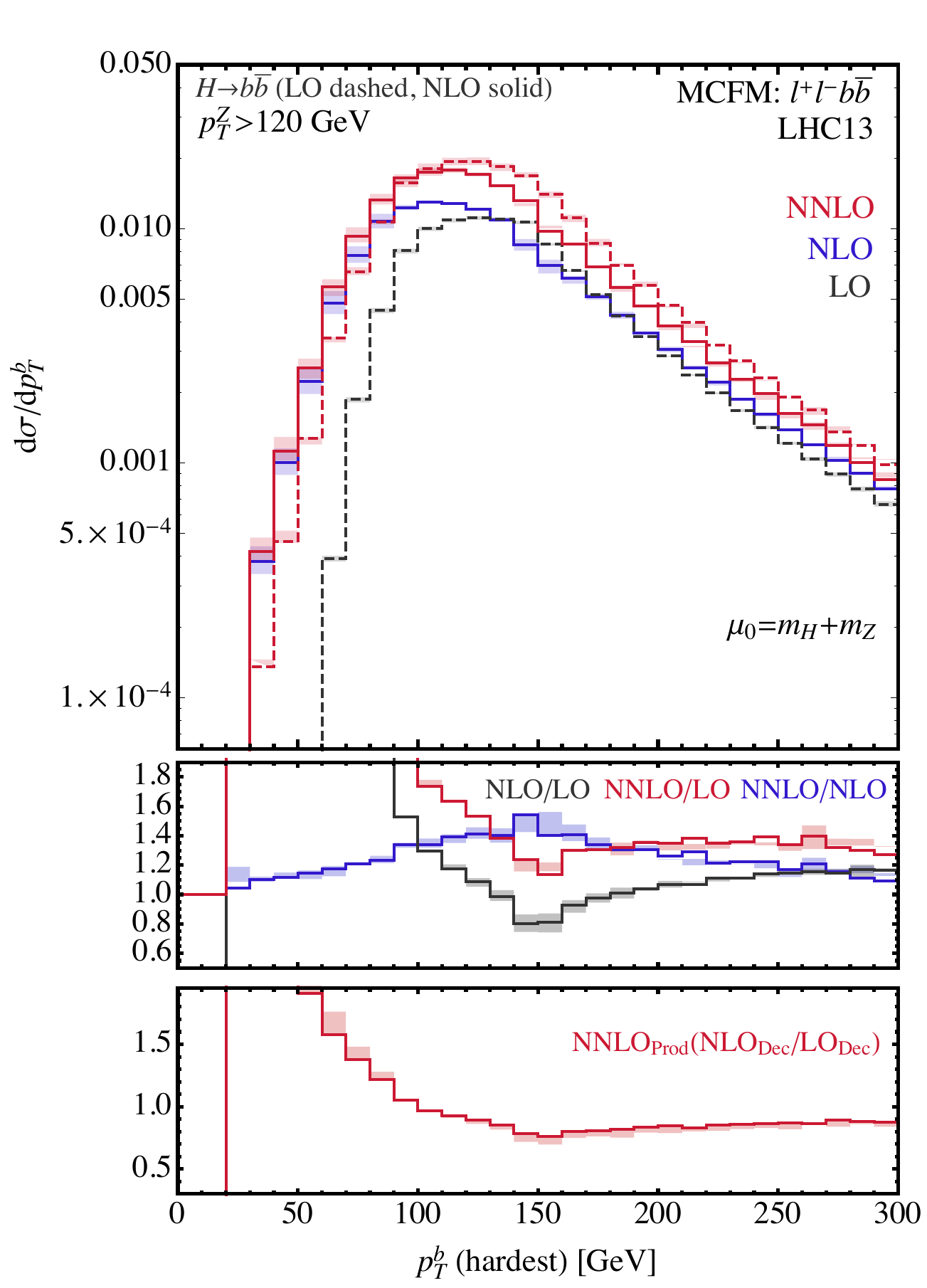} 
\includegraphics[width=0.48\textwidth]{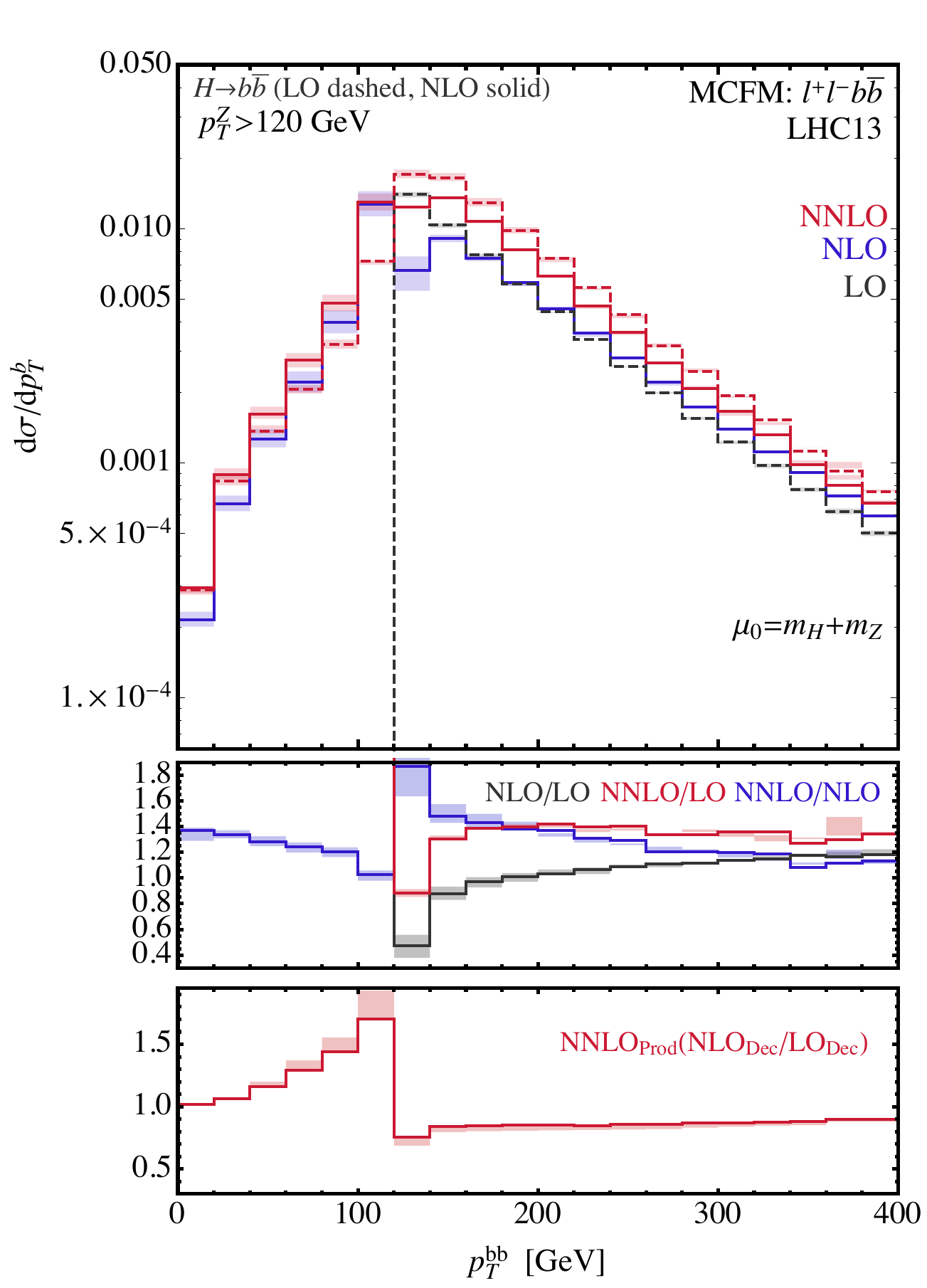} 
\caption{Differential predictions for the transverse momenta of the hardest $b$ (left) and the $b\overline{b}$ system (right) for $ZH$ at the LHC. Predictions at NNLO in production plus NLO decay 
are denoted by the solid lines, while NNLO in production plus LO decays are shown with dashed curves. }
\label{fig:Zbplots}
\end{center} 
\end{figure}

The impact of including the NLO decay is shown differentially in
Figs.~\ref{fig:Wbplots} $(WH)$ and Figs.~\ref{fig:Zbplots} $(ZH)$. We
present differential distributions for the hardest $b$-jet (left) and
$p_{T}^{b\overline{b}}$ (right), applying the basic lepton cuts,
demanding two $b$-jets and enforcing $p_T^V > 120$~GeV.  The general impact of
including the higher order corrections in the decay is immediately
apparent as a general softening of both spectra. This is easily
understood from the decay kinematics.   The invariant mass of the
system is constrained to be very close to $m_H^2$. When there are three particles
present in the decay to share the energy, the result is a softer
spectrum. For $p_{T}^{b\overline{b}}$ the impact of higher order
corrections in production and decay are particularly important. At LO,
$p_{T}^{b\overline{b}}=p_{T}^{V}$ so that the cut on the vector
boson momentum is also a cut on $p_{T}^{b\overline{b}}$. At NLO this is
no longer necessarily the case, since the real corrections allow for
an unclustered parton to balance the total momentum. Therefore the
region $p_{T}^{b\overline{b}} < 120$~GeV is first accessible at
NLO. Since in this region of phase space the total transverse momentum
of the $b\overline{b}$ is by definition relatively small, the resulting
transverse momentum of the $b$-quark pair is also relatively soft. As
a result the region of phase space where $p_T^b$(hard)~$<120$~GeV also
has large higher order corrections. This is highlighted in the middle
panel of the figures which presents the impact of higher order
corrections in production (for NLO decays).  Going from LO to NLO
there are large corrections to the $p_T$ spectrum of the
hardest $b$ quark, however the NNLO prediction is relatively stable
illustrating that the perturbative expansion is well-behaved beyond
LO.

For $p_T^{b\overline{b}}$ there is a strong feature
at the edge of the phase space for the NLO decays that is not present for the
LO decay option. This is due to the phase space boundary at the LO threshold in the decay phase space~\cite{Frixione:1997ks}.
The virtual decay corrections reside in the $p_{T}^{b\overline{b}} > 120$ GeV region of phase space, whilst the real 
corrections $H\rightarrow b\overline{b} g$ can 
fill the region both above and below this value. However in the real phase space when $p_{T}^{b\overline{b}} = 120$~GeV
there is a restriction on the phase space  for soft gluon emission and a large logarithm arises. 
Boundary problems such as these occur frequently in perturbation theory~\cite{Frixione:1997ks} and have been observed for
this specific process in previous calculations~\cite{Ferrera:2013yga,Ferrera:2014lca}.
For $ZH$ production the spectrum is smoothed-out somewhat by the turn-on of the $gg\rightarrow ZH$ contribution,
which resides  in the Born phase space.

\subsection{Results: $H\rightarrow WW^*\rightarrow \ell^+\nu\ell^-\overline{\nu}$}

The decay $H\rightarrow
WW^*$ represents a significant fraction, about $20\%$, of the total decay rate. Although requiring
leptonic decays of the $W$ bosons reduces the rate further, the signal cross
sections are large enough to have warranted experimental investigation in Run I
of the LHC~\cite{Aad:2015ona}. Going forward into Run II, triboson signatures represent a
fresh environment in which to search for new physics. From the technical point of view the $2\rightarrow 8$  phase
space is large, corresponding to a 22-dimensional Monte Carlo integration in the double-real part
of the calculation. This therefore represents a
demanding application of the $N$-jettiness slicing technique that tests its
suitability for complex phenomenological studies at NNLO. 
We study the production of $VH(\rightarrow WW^*)$ at the 14 TeV LHC with the following simple
phase space selection criteria,
\begin{eqnarray} 
{\rm{Jets : }}&& \quad  p_T^j > 25 \; {\rm{GeV}}, \;  |\eta_{j} | < 2.5  \\
{\rm{Leptons :}}&& \quad p_T^{\ell} > 25 \; {\rm{GeV}}, \; |\eta_{\ell} | < 2.5 \\
{\rm{MET: }} &&  \quad \slashed{E}_T > 20 \;  {\rm{GeV}}
\end{eqnarray}
Results for the cross-sections obtained under these cuts are presented in Table~\ref{table:VHWW}.
We present results for a single family of leptons for each  $V$ decay. Including all possibilities
of electron and muon configurations would thus increase the total rates presented in the table by
a factor of eight. We do not consider any interference between the  leptons arising from the decay of
the associated vector boson and those arising from the Higgs boson decay. The primary aim of this section is to address the
feasibility of producing NNLO results using the $N$-jettiness slicing method for a high-dimensional final
state configuration. We therefore postpone the treatment of interference effects for a future
study.
\begin{table}
\begin{center}
\begin{tabular}{|c|c|c|c|}
\hline
Process & $\sigma^{VH}_{LO} $ [fb] & $\sigma^{VH}_{NLO}$ [fb] & $\sigma^{VH}_{NNLO}$ [fb]  \\
\hline\hline 
$W^+H\rightarrow \ell_1^+\ell^+_{2}\ell_3^- +\slashed{E}_T$   & 0.0293$^{+0.9\%}_{-3.8\%}$    &   0.0394$_{-3.1\%}^{+2.3\%}$ & 0.0412$^{+1.6\%}_{-0.9\%}$   \\
\hline
$W^-H\rightarrow \ell_1^-\ell^+_{2}\ell_3^- +\slashed{E}_T$     &  0.0180$^{+3.0\%}_{-3.8\%}$ &  0.0250$_{-2.4\%}^{+1.6\%}$ & 0.0261$^{+0.6\%}_{-0.5\%}$   \\
\hline
$ZH\rightarrow \ell_1^+\ell_1^- \ell^+_2 \ell_3^- +\slashed{E}_T$ &  0.00634$_{-3.3\%}^{+2.4\%}$   & 0.00854$_{-1.9\%}^{+2.7\%}$   &  0.0104$^{+2.1\%}_{-2.1\%}$    \\
\hline 
\end{tabular}
\caption{Cross-sections for $VH \rightarrow VWW^* \rightarrow $ leptons at the 14 TeV LHC.
Results are presented for a single family of leptons.}
\label{table:VHWW}
\end{center}
\end{table}

An important variable when considering the decay $H \to WW^*$ is the transverse mass
of the electroweak final state.  For the $WH$ process it is defined by,
\begin{eqnarray} 
m^{WH}_T = \sqrt{(E_T^{3\ell}+\slashed{E}_T)^2-|{\bf{p}}_T^{3\ell}+{\bf{\slashed{E}_T}}|^2}
\end{eqnarray}
where 
\begin{eqnarray}
E_T^{3\ell} = \sqrt{{|{\bf{p}}_T^{3\ell}|}^2+m_{3\ell}^2} \;.
\end{eqnarray}
The equivalent definition for $ZH$ production ($m^{ZH}_T$) is obtained by making
the replacement $3\ell\rightarrow 4\ell$. 
The transverse mass is important since it can be used as a proxy for $m_{VH}$, which
is not experimentally observable for $H\rightarrow WW^*$ decays. 
The study of this variable is interesting due to its sensitivity to high-energy
structures that may be present in the $HVV$ vertex in BSM scenarios.  For instance,
treating the SM as an EFT introduces six- (and higher-) dimensional operators that induce
momentum dependent couplings between the Higgs and the vector bosons\footnote{A study of these operators 
at NLO (+PS) in the MCFM framework was presented recently in ref.~\cite{Mimasu:2015nqa}.}.  In general these operators
will induce small deviations in the tails of the transverse mass distribution, so it is crucial to have
control over percent-level effects from the SM in this region. 

\begin{figure}
\begin{center} 
\includegraphics[width=0.48\textwidth]{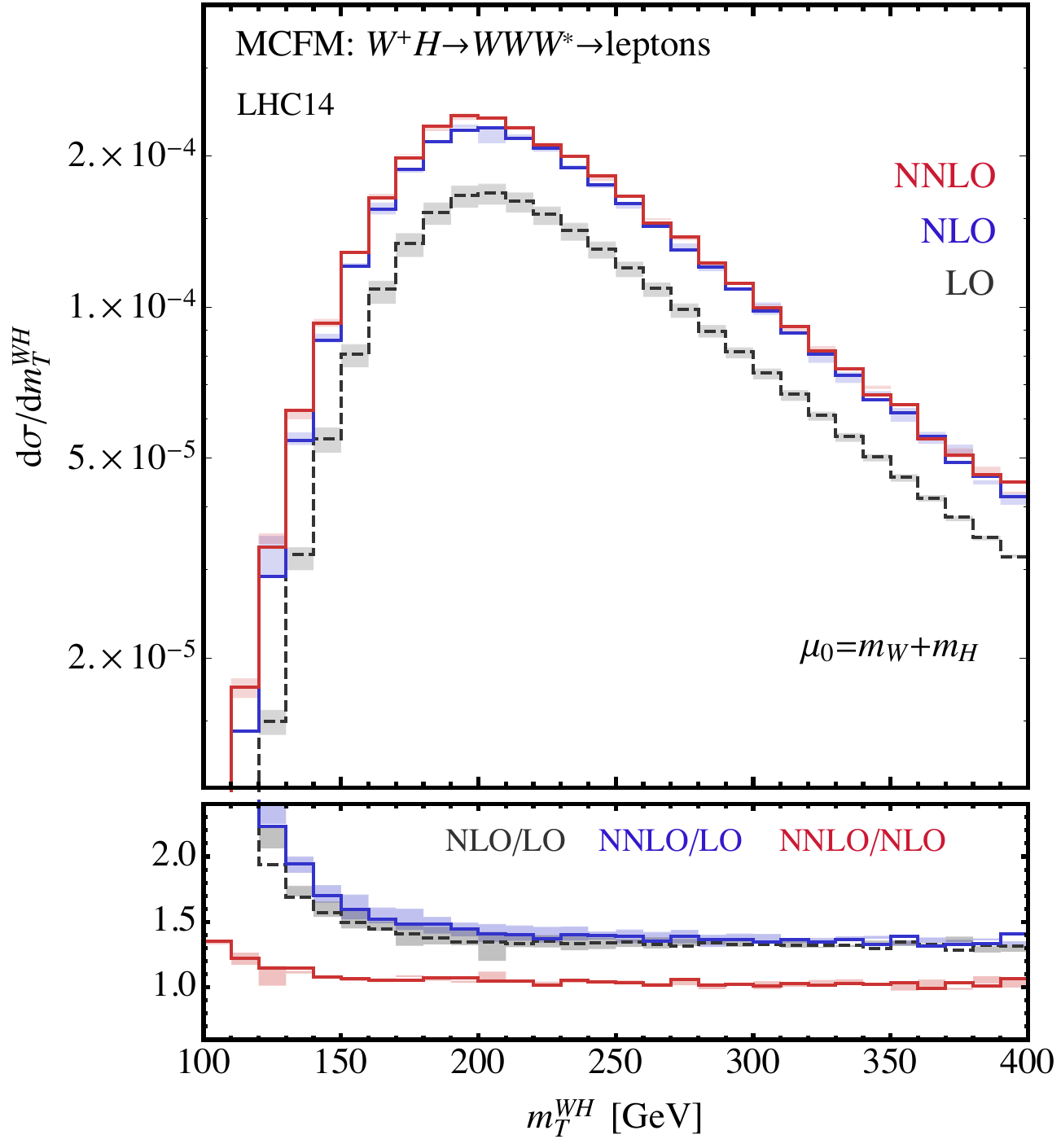} 
\includegraphics[width=0.48\textwidth]{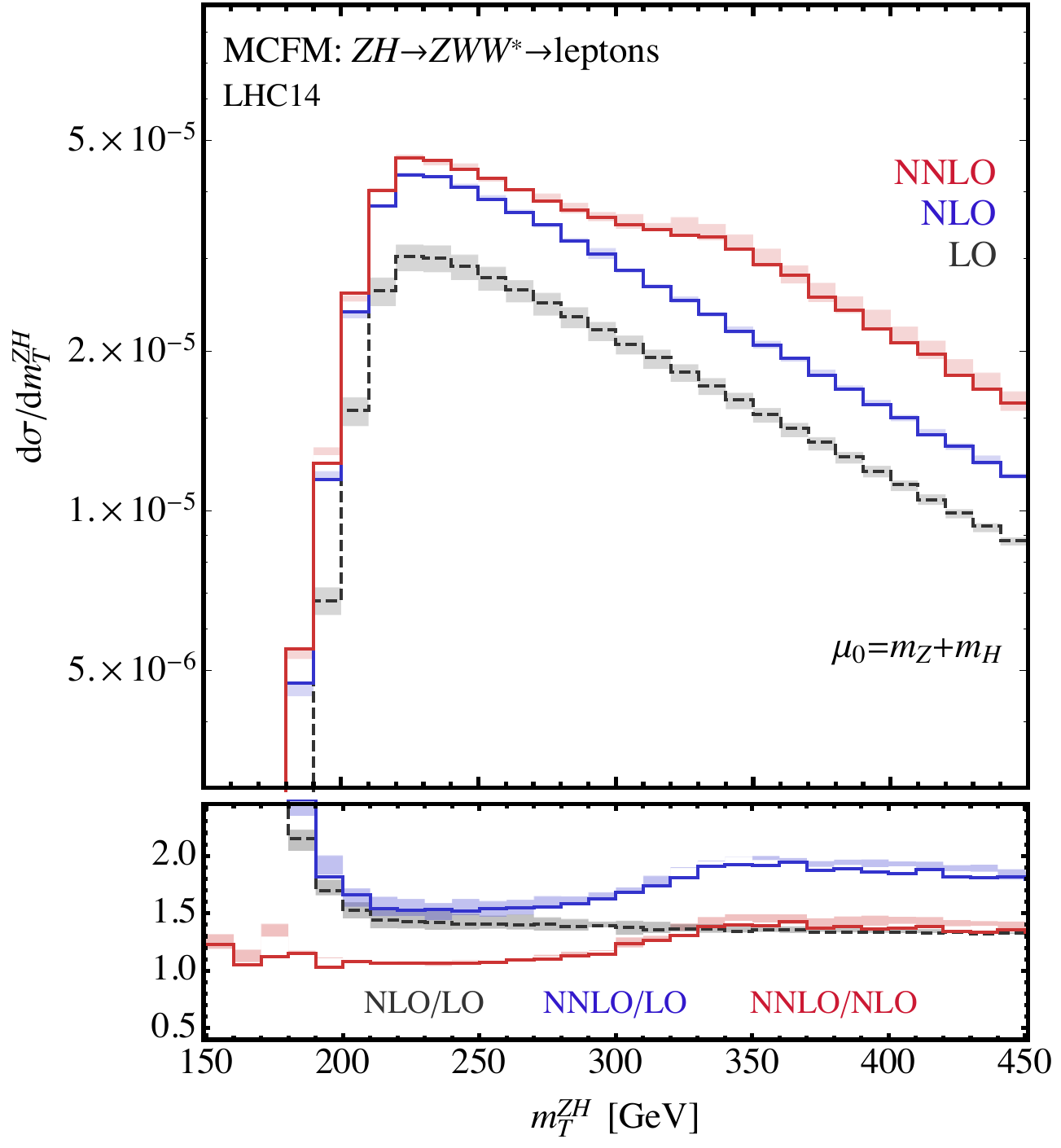} 
\caption{Differential predictions for the transverse mass of the lepton-$\slashed{E}_T$ system for $W^+H$(left) and
$ZH$ (right) production at the 14 TeV LHC. }
\label{fig:mtransvh}
\end{center} 
\end{figure}

Our predictions for this observable are presented in Fig.~\ref{fig:mtransvh} for $W^+H$ (left) and
$ZH$ (right). 
%We note that, although the cross sections presented in Table~\ref{table:VHWW} are obtained with $\tau^{\mbox{\tiny{cut}}}=0.01$~GeV, we present differential 
%distributions obtained with a slightly larger value, $\tau^{\mbox{\tiny{cut}}}=0.04$~GeV.  This choice is made in order to obtain smoother predictions for distributions and,
%given the results of section~\ref{sec:Hbb} and ref.~\cite{Campbell:2016yrh}, we expect any difference to be negligible. 
%From Fig.~\ref{fig:mtransvh} 
The difference in shape between $m^{WH}_T$  ($W^+H$ left) and $m^{ZH}_T$   ($ZH$ right) is apparent. Since
the transverse mass for $ZH$ is closer in definition
to $m_{VH}$, it has a
much harder spectrum, whereas $m^{WH}_T$ is much softer. The higher-order
corrections for $ZH$ are much larger, primarily due to the significant $gg\rightarrow ZH$ contribution.
These pieces 
are sensitive to the $2 m_{t}$ thresholds present in the loop integrals, and as a result turn on
at $m^{ZH}_T \sim 300$ GeV.  In this region the NNLO corrections are much larger
and the shape of $m^{ZH}_T$ is significantly altered. It is therefore essential
to include NNLO predictions for this observable in order to avoid attributing any observed change
in shape to the presence of BSM physics. 

\begin{figure}
\begin{center} 
\includegraphics[width=0.48\textwidth]{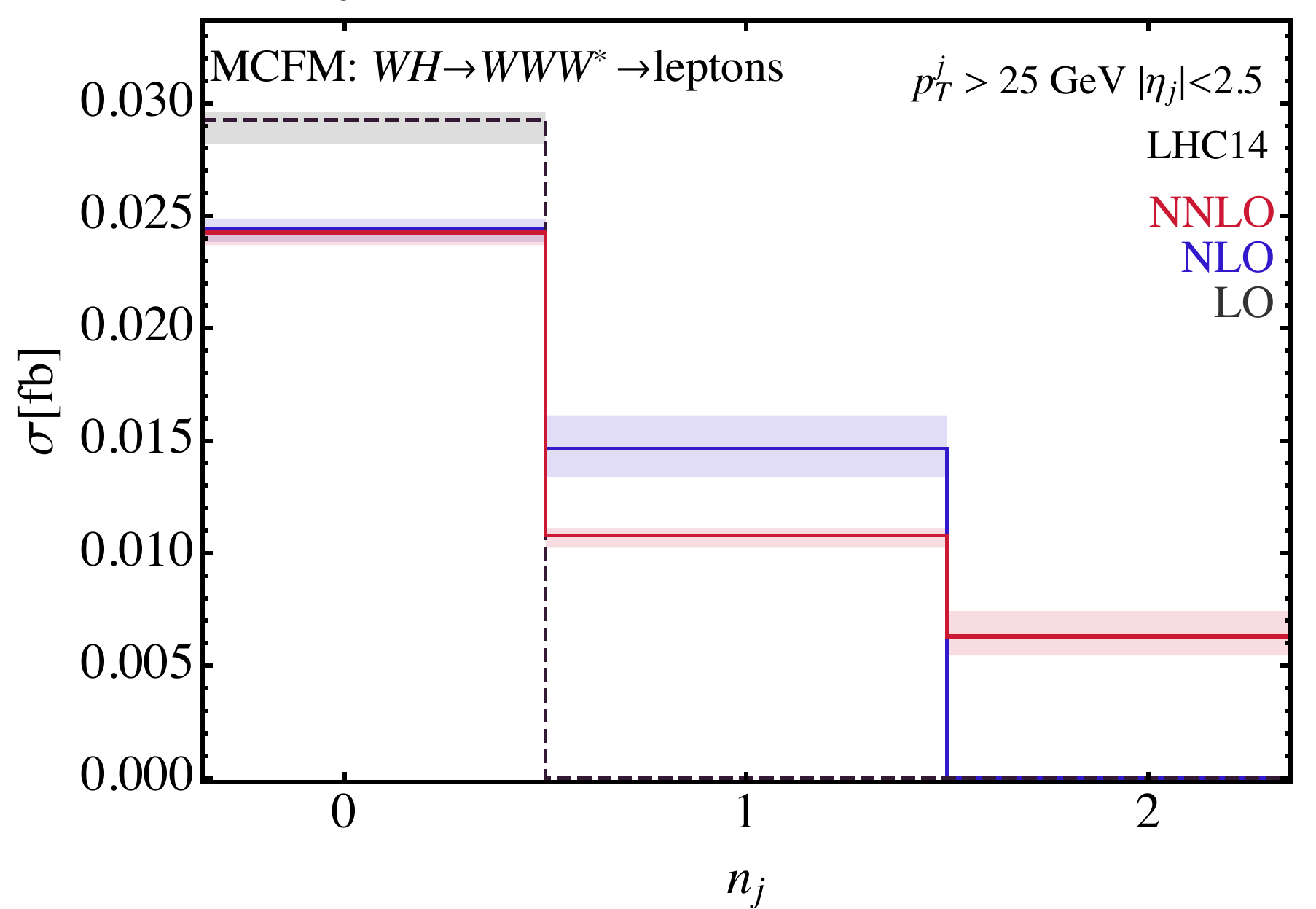} 
\includegraphics[width=0.48\textwidth]{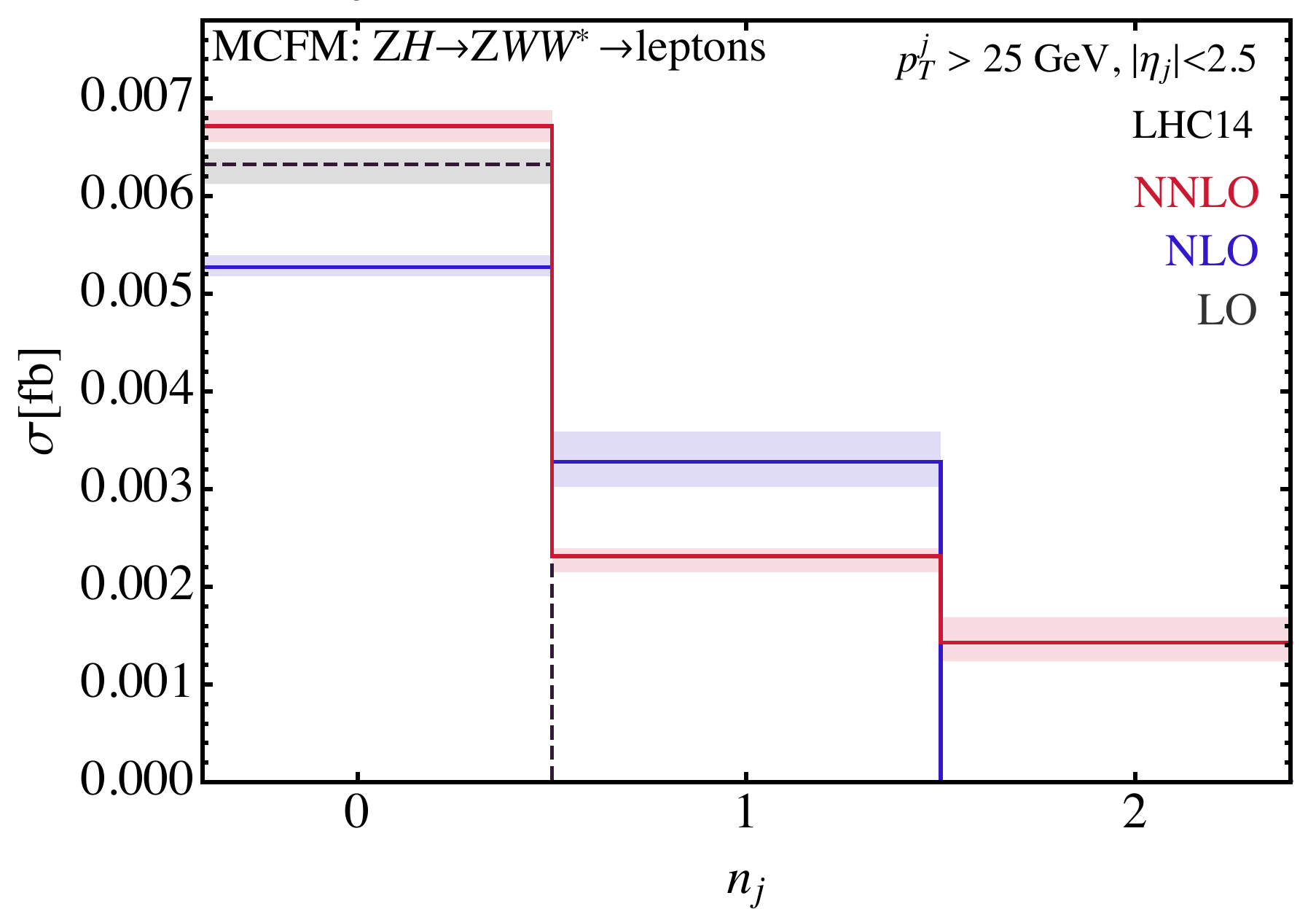} 
\caption{The cross-section as a function of the number of jets, $n_{j}$, for $W^+H$ (left) and $ZH$ (right) with $H\rightarrow WW^*$ decays at the 14 TeV LHC}
\label{fig:njHWW}
\end{center} 
\end{figure}

Finally in Fig.~\ref{fig:njHWW} we present the cross-section as a function of the number of additional jets,
where the basic jet definition is used from the previous section, $p_T^j > 25$ GeV and $|\eta_{j}| < 2.5$. The
$n_{j}$ distribution for these decays are different from those studied previously in the $H\rightarrow
b\overline{b}$ section, since now the jets are only produced through initial state radiation, with no
contamination from jets arising from the decay. For the $WH$ process, around 40\% of the events have one or
more jets in the final state. For $ZH$ production the percentage drops to around 35\% due to the presence of the
$gg\rightarrow ZH$ contribution that only populates the 0-jet bin.

\subsection{Results: $H\rightarrow \gamma\gamma$}

\begin{figure}
\begin{center} 
\includegraphics[width=0.48\textwidth]{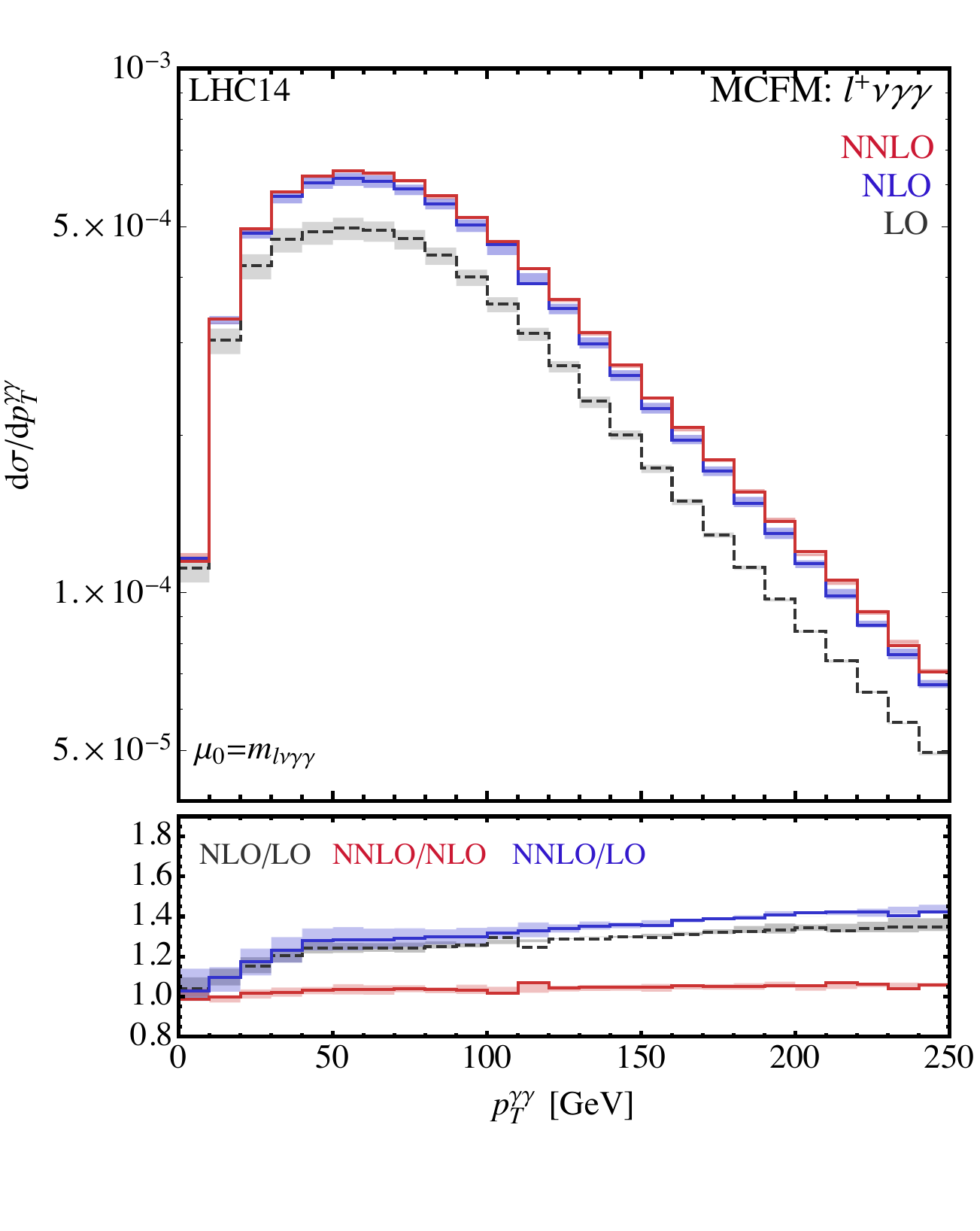} 
\includegraphics[width=0.48\textwidth]{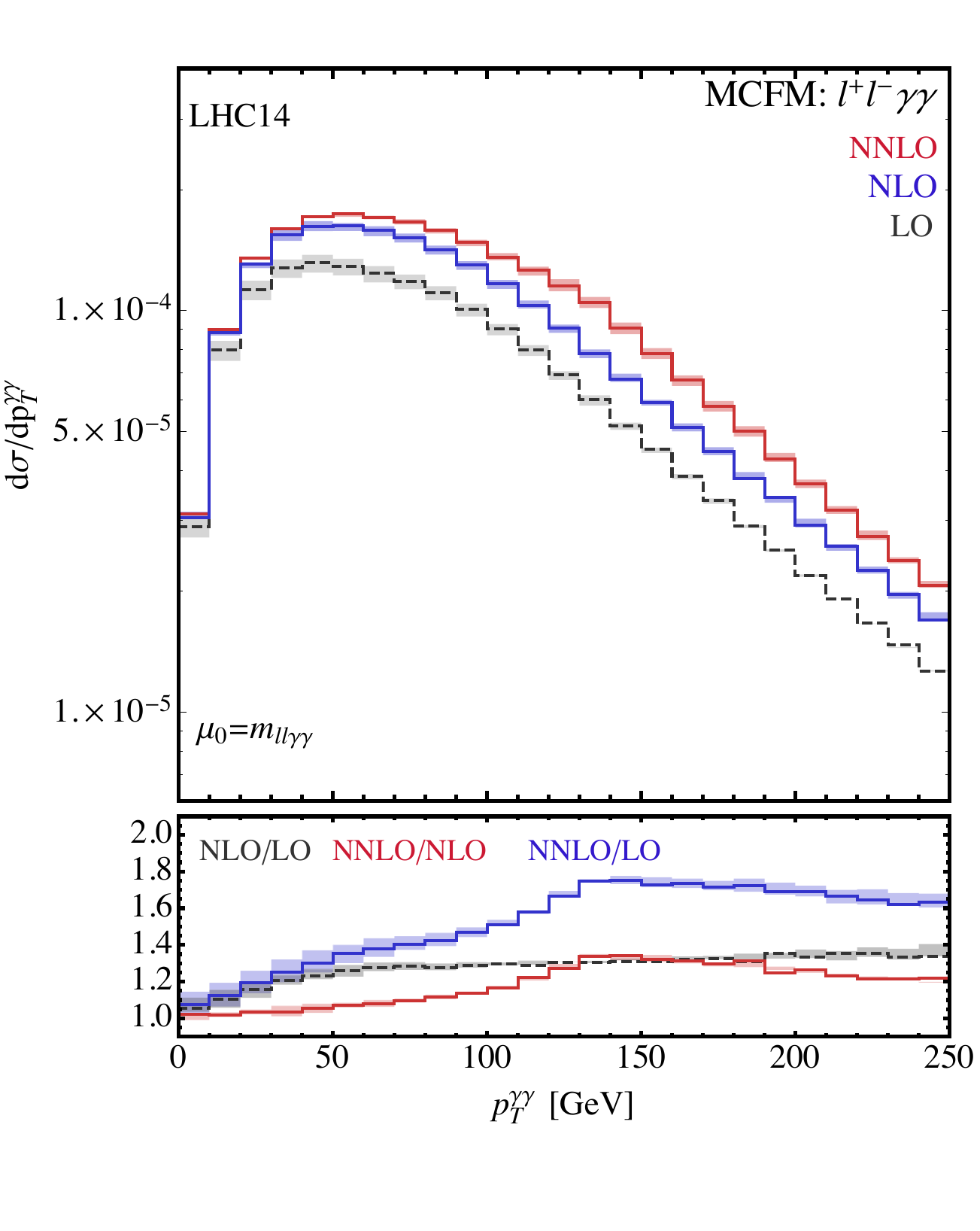} 
\caption{Differential predictions for the transverse momenta of the photon pair for $W^+H$(left) and  $ZH$ (right) at the LHC. }
\label{fig:ptgaga}
\end{center} 
\end{figure}

% \begin{table}
% \begin{center}
% \begin{tabular}{|c|c|c|c|c|}
% \hline
% Process & $\sigma^{VH}_{LO} $ [fb] & $\sigma^{VH}_{NLO}$ [fb] & $\sigma^{VH}_{NNLO}$ [fb] &$\sigma^{\rm{Back}}_{LO,m_{\gamma\gamma}}$ \\
% \hline\hline 
% $\ell^+\nu\gamma\gamma$   & 0.0686$^{+3.0\%}_{-3.7\%}$  &  0.0860$^{+2.2\%}_{-2.4\%}$   &  0.0886$^{+0.2\%}_{-0.8\%}$ &   0.04586$^{+1.9\%}_{-2.4\%}$   \\
% \hline
% $\ell^-\overline{\nu}\gamma\gamma$  & 0.0448$^{+3.2\%}_{-3.9\%}$  &  0.0579$^{+2.3\%}_{-2.4\%}$  & 0.0599$^{+0.3\%}_{-0.7\%}$ & 0.0324$^{+2.7\%}_{-3.9\%}$ \\
% \hline
% $ \ell^+\ell^-\gamma\gamma $ & 0.0177$^{+3.1\%}_{-3.9\%}$    &  0.02243$^{+2.2\%}_{-2.4\%}$ &  0.0256$^{+1.4\%}_{-1.9\%}$ & 0.0555$^{+1.6\%}_{-4.6\%}$    \\
% \hline 
% \end{tabular}
%%%%%
%%%%% John's updated NNLO numbers below
%%%%%
\begin{table}
\begin{center}
\vskip 0.5cm
\begin{tabular}{|c|c|c|c|c|}
\hline
Process & $\sigma^{VH}_{LO} $ [fb] & $\sigma^{VH}_{NLO}$ [fb] & $\sigma^{VH}_{NNLO}$ [fb] &$\sigma^{\rm{Back}}_{LO,m_{\gamma\gamma}}$ \\
\hline\hline 
$\ell^+\nu\gamma\gamma$   & 0.0686$^{+3.0\%}_{-3.7\%}$  &  0.0860$^{+2.2\%}_{-2.4\%}$  
% & $0.0891+0.0002=0.0893 , 0.0872+0.0019=0.0891, 0.0852+0.0032=0.0884$ %%% new NNLO numbers here (up, central, down)
 & $0.0891^{+0.2\%}_{-0.8\%}$ %%% new NNLO numbers here
 & 0.0459$^{+1.9\%}_{-2.4\%}$   \\
\hline
$\ell^-\overline{\nu}\gamma\gamma$  & 0.0448$^{+3.2\%}_{-3.9\%}$  &  0.0579$^{+2.3\%}_{-2.4\%}$ 
% & $0.0604+0.0000=0.0604, 0.0591+0.0012=0.0603, 0.0577+0.0020=0.0597$ %%% new NNLO numbers here (up, central, down)
 & $0.0603^{+0.2\%}_{-1.0\%}$ %%% new NNLO numbers here
 & 0.0324$^{+2.7\%}_{-3.9\%}$ \\
\hline
$ \ell^+\ell^-\gamma\gamma $ & 0.0177$^{+3.1\%}_{-3.9\%}$    &  0.0224$^{+2.2\%}_{-2.4\%}$
% & $0.0233+0.0028=0.0261, 0.0228+0.0028=0.0256, 0.0223+0.0029=0.0252$ %%% new NNLO numbers here (up, central, down)
 & $0.0256^{+2.0\%}_{-1.6\%}$ %%% new NNLO numbers here
 & 0.0555$^{+1.6\%}_{-4.6\%}$    \\
\hline 
\end{tabular}
%%%
%%%
%%%
\caption{Cross-sections for $V\gamma\gamma$ production at the 14 TeV LHC, for a single family of leptons.
% using the central scale choice $\mu_0=m_{\ell\ell\gamma\gamma}$.
The corresponding phase space selection criteria are described in the text.}
\label{table:VHgaga}
\end{center}
\end{table}
The decay $H\rightarrow \gamma\gamma$ provides a relatively clean experimental signature, at the cost of a very small branching ratio. However 
during Run II enough data should be collected to allow experimental studies of this channel. 
In Table~\ref{table:VHgaga} we collect cross-sections for $VH(\to \gamma\gamma)$ processes at the LHC
operating at 14 TeV, after application of the following basic selection criteria:
\begin{eqnarray} 
{\rm{Photons : }}&& \quad  p_T^{\gamma_1} > 40 \; {\rm{GeV}}, \; p_T^{\gamma_2} > 25 \; {\rm{GeV}}, \;  |\eta_{\gamma} | < 2.5  \nonumber\\
&&  \quad R_{\gamma\gamma} > 0.4 \quad R_{\gamma j} > 0.4 \quad  R_{\gamma \ell} > 0.4 \ \\
{\rm{Jets : }}&& \quad  p_T^j > 25 \; {\rm{GeV}}, \;  |\eta_{j} | < 2.5  \\
{\rm{Leptons :}}&& \quad p_T^{\ell} > 25 \; {\rm{GeV}}, \; |\eta_{\ell} | < 2.5 \\
{\rm{MET: }} &&  \quad \slashed{E}_T > 20 \;  {\rm{GeV}}
\end{eqnarray}
We set the central renormalization  and factorization scale as $\mu_0 = m_{\ell_1\ell_2\gamma\gamma}$ and, as
in the previous section, vary the central scale by a factor  of two in opposite directions.
Although the cross-sections for this process are
rather small,  around $0.05$ fb, the advantage they possess over $H\rightarrow b\overline{b}$ decays is a much
smaller irreducible background. We illustrate this by including in 
Table~\ref{table:VHgaga} the background cross-sections in the Higgs resonance region
($\sigma^{\rm{Back}}_{LO,m_{\gamma\gamma}}$) that is defined by,
\begin{equation} 
120\;{\rm{GeV}} < m_{\gamma\gamma} < 130\; {\rm{GeV}}
\end{equation}
These cross-sections are for illustration only and are therefore evaluated at LO. The results of
Table~\ref{table:VHgaga} clearly demonstrate the potential for an excellent signal-to-background ratio in
this channel, although we note that experimental analyses would also have to cope with a large reducible background from
$V$+jets that may contaminate this significantly.  Unsurprisingly the impact of the NNLO corrections for this decay channel are essentially the
same as those discussed in detail in the previous sections. We demonstrate the impact on a canonical differential 
observables in Fig.~\ref{fig:ptgaga}, which depicts the $p_T^{\gamma\gamma}$ spectrum for both $W^+H$ and $ZH$
production at the 14 TeV LHC. 

\section{Conclusions} 
\label{sec:conc}

In this paper we have presented a NNLO calculation of $VH$ production and its implementation into a fully flexible Monte Carlo code. 
These processes can provide a useful handle on the coupling of the Higgs boson to bottom quarks, in addition to serving as sensitive
probes of anomalous interactions between the Higgs, $W$ and $Z$ bosons.
At NNLO the calculation of these processes includes both Drell-Yan-like contributions and corrections where the Higgs boson is radiated from a heavy quark
loop rather than from the vector boson.  These two contributions are comparable in size for $WH$ production.  For the $ZH$ process
the latter corrections involve $gg$-initiated diagrams that dominate the NNLO correction.  Including both contributions is
therefore imperative and our calculation enables the combined effects to be studied in a differential manner for the
first time.   Analytic results for these amplitudes can be found in the appendix and the distributed MCFM code.  

The $VH$ processes are among the most interesting Higgs production modes for phenomenological studies in Run II.
We have therefore studied a number of decay modes of the Higgs boson in some detail.  In the case of the
decay to a pair of bottom quarks, $H\rightarrow b\overline{b}$, we have also included the effect of radiation in the 
decay at NLO accuracy.  This has a considerable impact on, for instance, the transverse momentum of the $b\bar b$ pair.
Our results suggest that including QCD corrections in the decay to NNLO is important, although it is beyond the scope
of this paper.  We have investigated the phenomenology of other Higgs decay channels that might be explored more fully
in Run II of the LHC, namely the decays  $H\rightarrow WW^* \rightarrow$ leptons and $H\rightarrow\gamma\gamma$.

From the theoretical point of view, presenting a NNLO
calculation for the $2\rightarrow 6$ process, $pp \to VH \to VWW^* \to$~leptons+$\slashed{E}_T$ is technically
challenging due to the large final state phase space. In the double-real part
of the calculation the phase space is 22-dimensional, and provides a challenging environment in which to test the jettiness-based
approach to NNLO calculations.  The $H\rightarrow \gamma\gamma$ decay results in very small cross-sections, but has the advantage
that the irreducible background in the neighborhood of the Higgs boson mass is also small, resulting in comparable signal and
background rates.

The results of this paper have been implemented into MCFM, and are publicly available.

\acknowledgments

We thank Radja Boughezal, Xiaohui Liu and Frank Petriello for assisting in the implementation
of the $N$-jettiness regularization procedure in MCFM. 
CW thanks Shawn Matott for computational help and JC is grateful to the Fermilab computing sector for
providing access to the Accelerator Simulations Cluster.
Fermilab is supported by the US DOE under contract DE-AC02-07CH11359. Support provided by the Center for Computational Research at the University at Buffalo.

\appendix

\section{Amplitudes for DY contributions}
\label{app:DYamp}

In this appendix we present analytic expressions for the parts of the calculation that are
closely-related to the Drell-Yan process. We present results for  contributions above and below
$\tau^{\mbox{\tiny{cut}}}$ separately. Since the Higgs boson is a scalar particle the decay amplitude
factorizes from  the production amplitude, modulo ${\cal O}(\alpha_s^2)$ corrections that we
neglect in our calculation. Therefore the results in this section are presented for an on-shell
Higgs boson and a $V$ boson that decays to leptons are included.  We have used the Mathematica
package S@M~\cite{Maitre:2007jq} frequently in our calculations.

\subsection{Below $\tau^{\mbox{\tiny{cut}}}$}

Below $\tau^{\mbox{\tiny{cut}}}$ the calculation requires the soft, beam, and hard functions of the SCET
formalism.  The necessary two-loop soft and beam functions were computed in 
refs.~\cite{Kelley:2011ng,Monni:2011gb} and~\cite{Gaunt:2014xga} respectively. The process-dependent  hard-function can be extracted from the two-loop virtual 
form-factor as in refs~\cite{Idilbi:2006dg,Becher:2006mr}. We have repeated the  calculation
and for completeness we reproduce the $1$- and $2-$loop results below, 

\begin{eqnarray}
\left|\mathcal{M}^{(1)}\right\rangle_{\overline{MS}} &=& C_F\left(\frac{\as}{2\pi}\right) \left(-L^2+3L-8+\zeta_2\right)|\mathcal{M}^{(0)}\rangle \\
\left|\mathcal{M}^{(2)}\right\rangle_{\overline{MS}} &=&\left(\frac{\as}{2\pi}\right)^2 \bigg[C_F^2\bigg(\frac{1}{2}(L^2-3L+8-\zeta_2)^2 \nonumber\\&&
      +\left(\frac{3}{2}-12\zeta_2+24\zeta_3\right)L 
      -\frac{1}{8}+29\zeta_2-30\zeta_3-\frac{44}{5}\zeta_2^2\bigg) \nonumber\\&&
      +C_FN_f\bigg(-\frac{2}{9}L^3+\frac{19}{9}L^2-\left(\frac{209}{27}+\frac{4}{3}\zeta_2\right)L
     +\frac{4085}{324}+\frac{23}{9}\zeta_2+\frac{2}{9}\zeta_3\bigg)\nonumber\\&&
     +C_FC_A\bigg(\frac{11}{9}L^3+\left(2\zeta_2-\frac{233}{18}\right)L^2
      +\left(\frac{2545}{54}+\frac{22}{3}\zeta_2-26\zeta_3\right)L\nonumber\\&&
      -\frac{51157}{648}-\zeta_2\left(\frac{337}{18}-\frac{44}{5}\zeta_2\right)+\frac{313}{9}\zeta_3\bigg)\bigg]
      \left|\mathcal{M}^{(0)}\right\rangle
\end{eqnarray}
where $L=\log{\left({-s_{12}}/{\mu^2}\right)}$ and $\left|M^{(0)}\right\rangle$
represents the LO amplitude. Note that the above amplitudes are defined in four-dimensions
and as such the LO amplitude (for $WH$ production) can be defined in terms of helicity
amplitudes as follows, 
\begin{eqnarray} 
\left|\mathcal{M}^{(0)}\right\rangle_{WH}=g_W^3 m_W \mathcal{P}_W(s_{34})
\mathcal{A}^{(0)}(1^-_{q},2^+_{\overline{q}},3^+_{\overline{\ell}},4^-_{\ell},p_H)
\end{eqnarray} 
In the above equation $\mathcal{P}$ represents the propagator function, 
\begin{eqnarray}
\mathcal{P}_X(s) = \frac{s}{s-m_X^2+i m_X\Gamma_X} 
\end{eqnarray}
The tree-level helicity amplitude is then defined as follows\footnote{We refer readers unfamiliar with spinor-helicity notation to one of the many 
comprehensive reviews, for instance ref.~\cite{Dixon:1996wi}}, 
\begin{eqnarray}
\mathcal{A}^{(0)}(1^-_{q},2^+_{\overline{q}},3^+_{\overline{\ell}},4^-_{\ell},p_H)= \frac{\spa1.4\spb 3.2}{s_{34}s_{12}}
\end{eqnarray}
It is instructive to re-write this amplitude as, 
\begin{eqnarray}
\mathcal{A}^{(0)}(1^-_{q},2^+_{\overline{q}},3^+_{\overline{\ell}},4^-_{\ell},p_H)= -\frac{1}{s_{12}}\left(\frac{\spa 1.4^2}{\spa 1.2\spa3.4}+\frac{\spa1.4\spab 1.p_H.3}{\spa1.2 s_{34}}\right)
\end{eqnarray}
Modulo the overall factor of $s_{12}$, which is a result of the internal $W$ propagator function,
the holomorphic piece of the above expression (the first term) corresponds exactly to the  MHV
tree-level amplitude for $q\overline{q}\ell\overline{\nu}$. The second, non-holomorphic, term
is a correction that vanishes in the soft Higgs boson limit. Since the amplitudes for $W+3$
and $W+4$ partons have been calculated analytically~\cite{Giele:1991vf,Bern:1997sc} the above
decomposition provides a useful check of all of our calculated amplitudes.

The helicity breakdown for the $ZH$ process requires a summation over the left and right-handed couplings 
\begin{eqnarray} 
\left|\mathcal{M}^{(0)}\right\rangle_{ZH}=2\frac{g_W e^2}{\cos^2\theta_W}m_W \mathcal{P}_H(s_{1234})\mathcal{P}_Z(s_{34})
\sum_{ij = L,R} v^{q}_{i}v^{\ell}_{i} \mathcal{A}^{(0)}_{ij}
\end{eqnarray} 
The fermionic (quark or lepton) coupling to the $Z$ boson is given by $v_{h}^{(q,\ell)}$:
\begin{eqnarray}
v_{L}^{\ell}=\frac{-1-2Q_{\ell}\sin^2\theta_W}{\sin2\theta_W} \quad v_{R}^{\ell}=-\frac{2Q_{\ell}\sin^2\theta_W}{\sin2\theta_W} \nonumber\\
v_{L}^{q}=\frac{\pm1-2Q_{q}\sin^2\theta_W}{\sin2\theta_W} \quad v_{R}^{q}=-\frac{2Q_{q}\sin^2\theta_W}{\sin2\theta_W} 
\end{eqnarray}
The sign in the $v_{L}^{q}$ term is determined by whether the quark is up ($+$) or down ($-$)
type. The helicity amplitudes are then obtained from the equivalent amplitudes for the $WH$
process, by applying line reversal symmetries as necessary.

\subsection{Above $\tau^{\mbox{\tiny{cut}}}$}

Above $\tau^{\mbox{\tiny{cut}}}$ the calculation corresponds to a NLO one for the $VH +$ jet process. We have
calculated helicity amplitudes for this process which, to the best of our knowledge, have not been
presented in the literature before. The LO amplitude for $WHj$ can be written as follows, 
\begin{eqnarray} 
|\mathcal{M}^{(0)}\rangle_{WHj} =\sqrt{2}g_s g_W^3 m_W (T^{g_5})^{i_1}_{j_2} \;\mathcal{P}_H(s_{12345})\mathcal{P}_W(s_{34})
\sum_{h_5=\pm1}\mathcal{A}^{(0)}(1^+_{\overline{q}},2^-_q,3^-_{\ell},4^+_{\overline{\ell}},5_g^{h_5},p_H) \nonumber\\
\end{eqnarray} 
The tree-level MHV helicity amplitude is  defined as, 
\begin{eqnarray}
\mathcal{A}^{(0)}_4(1^-_{q},2^+_{\overline{q}},3^-_{\ell},4^+_{\overline{\ell}},5_g^{+},p_H)=-\frac{\spa1.3\spab1.P_{25}.4}{s_{125}s_{34}\spa1.5\spa2.5}
\label{eq:A0WHj}
\end{eqnarray}
where $P_{25}=p_2+p_5$.  The helicity amplitude for $h_5=-1$ can be readily obtained from the above by the conjugation operation. 
At NLO we require the one-loop amplitude for $WHj$ and the tree-level amplitudes for $WHjj$. The one-loop amplitude for $WHj$ can be written as follows, 
 \begin{eqnarray} 
|\mathcal{M}^{(1)}\rangle_{WHj} &=&\sqrt{2}g_s g_W^3 m_W N_c (T^{g_5})^{i_1}_{j_2} \;\mathcal{P}_H(s_{12345})\mathcal{P}_W(s_{34})
\\&\times&
\sum_{h_5=\pm1}\bigg(\mathcal{A}^{(1)}_1(1^-_{q},2^+_{\overline{q}},3^-_{\ell},4^+_{\overline{\ell}},5_g^{+},p_H) + \frac{1}{N_c^2}\mathcal{A}^{(1)}_2(1^-_{q},2^+_{\overline{q}},3^-_{\ell},4^+_{\overline{\ell}},5_g^{+},p_H)\bigg)
\nonumber
\end{eqnarray}   
As discussed in the previous subsection, the amplitudes for $VHj$ can be decomposed in terms of those for $Vj$ and a piece which vanishes in the soft limit (where it is understood that momentum conservation is altered accordingly in the $Vj$ amplitude). We use this decomposition on our one-loop amplitudes, defining, 
\begin{eqnarray}
\mathcal{A}^{(1)}_i
 =V_i  \mathcal{A}^{(0)}+F_i + S^H_i
\end{eqnarray}
where $\mathcal{A}^{(0)}$ is understood to be given by eq.~\eqref{eq:A0WHj}. $V_i$ and $F_i$ then correspond
exactly to the amplitudes obtained in the calculation of the $W+3$ parton one-loop amplitude, and $S_i^H$
corresponds to the missing piece which vanishes in the soft Higgs limit. For brevity we do not reproduce the results
for $V_i$ and $F_i$ here, they can be found in the literature~\cite{Bern:1997sc} or in the distributed MCFM code.
The leading colour contribution $S^H_1$ has the following form, 
\begin{eqnarray}
 S^H_1 = -\frac{3\spa2.1\spab3.P_{125}.4\spb5.2}{2s_{125}^2\spa2.5}L_0(-s_{25},-s_{125})
 \end{eqnarray}
while the subleading colour contribution $S^2_H$ is 
 \begin{eqnarray}
S^H_2&=&\frac{\spa1.2\spab 3.P_{125}.4 }{s_{125}\spa1.5^2}\FFom(-s_{12},-s_{25};-s_{125})         
 \nonumber\\&& +
\frac{\spa1.2\spb5.1\spab 3.P_{125}.4}{s_{125}\spa1.5}\left(    \frac{L_0(-s_{125},-s_{25})}{s_{25}}-\frac{L_0(-s_{125},-s_{12})}{s_{12}}\right)
 \nonumber\\&&-\frac{\spb5.1\spab 3.P_{125}.4}{s_{125}\spa1.5\spb2.1}
 \end{eqnarray}
 Here $\FFom$ represents the finite part of the one-mass box integral and is given by, 
 \begin{eqnarray}
 \FFom(s,t;P^2) = -2\left(\mathrm{Li}_2\bigg(1-\frac{P^2}{s}\bigg)+\mathrm{Li}_2\bigg(1-\frac{P^2}{t}\bigg)+\frac{1}{2}\log^2\left(\frac{s}{t}\right)
 +\frac{\pi^2}{6}\right)
\end{eqnarray} 
 and the auxiliary function $L_0(s,t)$ is,
 \begin{eqnarray}
 L_0(s,t)=\frac{\log{s/t}}{1-s/t}
 \end{eqnarray}
 The amplitudes for $ZH$ production can be obtained in similar fashion to the tree-level discussion in the previous
 sub-section, i.e. by appropriate dressing of the helicity amplitudes by the left- and right-handed couplings and
 modification of the electroweak pre-factor accordingly. 

\section{Amplitudes for $\mathcal{O}(\alpha_s^2)$ $y_t$ contributions} 

\label{app:ytamp}

In this section we present formulae for some of the amplitudes that contribute to the term
$d\sigma^{(2),y_t}$.  There are in principle five such terms, labelled in ref.~\cite{Brein:2011vx} as $V_I$, $V_{II}$,
$R_{I}$, $R_{II}$ and $gg$-initiated pieces.  In this section we follow closely the formalism and nomenclature used in that reference,
which presents these contributions for on-shell bosons.  Here we do not include any results for $R_{II}$ since its
effects are tiny and not accounted-for in our calculation.  We also do not present our analytic results for the
$gg \to ZH$ contribution, which are too lengthy to include here but may be inspected in the distributed MCFM code. 

\subsection{$V_I$ pieces}

We begin by considering the $V_I$ pieces, which occur for both $W$ and $Z$ associated production
(cf. Fig.~\ref{fig:WHtop}, right). Using the method of asymptotic 
expansions\footnote{see, for example, the discussion in refs.~\cite{Smirnov:1994tg,Harlander:1999cs}} it was shown in ref.~\cite{Brein:2011vx} that the leading 
terms in the expansion correspond to replacing the top quark loop by the effective $ggH$ vertex. Therefore we can
obtain $V_I$ by calculating  the results in the effective field theory. 
We write the one-loop amplitude as follows, 
\begin{eqnarray} 
|\mathcal{M}^{VI}\rangle_{WH}= -C_F g_w^2 \frac{\alpha_s^2}{6\pi^2 v}\mathcal{P}_{W}(s_{34}) \mathcal{A}^{(1)}_{VI}(1^+_{\overline{q}},2^-_q,3^-_{\ell},4^+_{\overline{\ell}},p_H)
\end{eqnarray}
where 
\begin{eqnarray} 
\mathcal{A}^{(1)}_{VI}(1^+_{\overline{q}},2^-_q,3^-_{\ell},4^+_{\overline{\ell}},p_H)&=&
\frac{1}{2}\left(\frac{\spa 2.3^2}{\spa 1.2 \spa 3.4} + \frac{\spb 4.1^2}{\spb 2.1\spb4.3}\right)\FFtme(-s_{134},-s_{234};-s_{34},-s_{1234}) \nonumber\\
&&+\left(\frac{\spa 2.3\spb 4.1}{2\spab 1.P_{34}.1} - \frac{\spa1.2\spa 3.4\spb 4.1^2}{4\spab 1.P_{34}.1^2} \right)\log{\left(\frac{-s_{34}}{-s_{134}}\right)}\nonumber\\
&&+\left(\frac{\spa2.3\spb4.1}{2\spab2.P_{34}.2}-\frac{\spa2.3^2\spb2.1\spb4.3}{4\spab2.P_{34}.2^2}\right)
\log{\left(\frac{-s_{34}}{-s_{234}}\right)}\nonumber\\
&&-\frac{\spa2.3^2\spb2.1}{4\spa3.4\spab2.P_{34}.2}-\frac{\spa1.2\spb4.1^2}{4\spab1.P_{34}.1\spb4.3}
\end{eqnarray}

The finite part of the two-mass easy box is defined as 
\begin{eqnarray}
\FFtme(s,t;P^2,Q^2)&=&-2\bigg[\mathrm{Li}_2\bigg(1-\frac{P^2}{s}\bigg)
+\mathrm{Li}_2\bigg(1-\frac{P^2}{t}\bigg)+\mathrm{Li}_2\bigg(1-\frac{Q^2}{s}\bigg)
\nonumber\\&&+\mathrm{Li}_2\bigg(1-\frac{Q^2}{t}\bigg)-\mathrm{Li}_2\bigg(1-\frac{P^2Q^2}{st}\bigg)
+\frac{1}{2}\ln^2\bigg(\frac{s}{t}\bigg)\bigg]
\end{eqnarray}

\subsection{$V_{II}$ pieces} 

Next we consider the $V_{II}$ diagrams, in which the vector boson also couples to the closed fermion loop.
(cf. Fig.~\ref{fig:ZHtop}, left).  These contributions therefore only exist for $ZH$ production.
The results of ref.~\cite{Brein:2011vx} show that the leading terms in the 
$m_t$ asymptotic expansion correspond to the $q\overline{q}ZH$  effective vertex multiplying a two-loop massless
tadpole diagram. The leading term in the expansion thus has the form of a tree-level amplitude 
\begin{eqnarray}
|\mathcal{M}^{V_{II}}\rangle_{ZH}=16\frac{\as}{4\pi^2}G_F e m_W (v_L^t-v_R^t)\sum_{i=L,R} 
v_i^{\ell}\mathcal{A}^{(0)}_{VII}(1^+_{\overline{q}},2^-_q,3^-_{\ell},4^+_{\overline{\ell}},p_H)
\end{eqnarray}
where 
\begin{eqnarray}
\mathcal{A}^{(0)}_{VII}(1^+_{\overline{q}},2^-_q,3^-_{\ell},4^+_{\overline{\ell}},p_H)
=\frac{\spb1.4\spa3.2}{s_{34}}
\end{eqnarray}
The $(v_L^t-v_R^t)$ factor arises since only the axial part of the two-loop massive tadpole contributes to the
amplitude. 

\subsection{$R_I$ pieces} 

Finally we present the results for the $R_I$ contribution (cf. Fig.~\ref{fig:WHtop}, left). These contributions are
universal and occur for either $WH$ or $ZH$ production. They correspond to one-loop calculations and can be
performed  either including the top quark mass in full, or in the effective field theory. The general structure of
the result is as follows, 
\begin{eqnarray}
|\mathcal{M}^{R_{I}}\rangle_{WH}= \frac{g_s g_w^2}{\sqrt{2}}\mathcal{P}_W(s_{34}) \mathcal{J}^{FT/EFT} \sum_{h_5=\pm 1} 
\mathcal{A}^{(0)}_{R_I}(1^-_{q},2^+_{\overline{q}},3^-_{\ell},4^+_{\overline{\ell}},5_g^{h_5},p_H)\mathcal{I}^{FT/EFT}
\end{eqnarray}
The helicity amplitude is universal,
\begin{eqnarray} 
\mathcal{A}^{(0)}_{R_I}(1^-_{q},2^+_{\overline{q}},3^-_{\ell},4^+_{\overline{\ell}},5_g^+,p_H)&=&
\frac{\spab3.P_{24}.5\spab 1.P_{234}.5\spb4.2}{s_{1234}s_{234}s_{34}}
       \nonumber\\&&  +\frac{\spa1.3\spb5.2
     (\spa 1.2\spb 4.1\spb5.2-\spa2.3\spb4.3\spb5.2
     +s_{134}\spb5.4)}{s_{1234}s_{134}s_{34}}
\end{eqnarray}
but the prefactor $\mathcal{J}$ depends on whether one is in the full or the effective field theory, 
\begin{eqnarray}
\mathcal{J}^{FT} &=& \frac{\as}{4\pi} \frac{m_t^2}{2m_W} g_w \\
\mathcal{J}^{EFT} &=& \frac{\as}{3\pi v} 
\end{eqnarray}
Finally, the function $\mathcal{I}$ is either equal to unity in the effective field theory or is a
combination of scalar loop integrals in the full theory,
\begin{eqnarray}
\mathcal{I}^{EFT}&=&1 \\
\mathcal{I}^{FT} &=& 8\bigg(\frac{1}{2}\left(1-\frac{4m_t^2}{s_{12345}-s_{1234}}\right)I_3(s_{12345},s_{1234},m_t^2)\\&&  
+\frac{s_{1234}}{(s_{12345}-s_{1234})^2}\left(I_2(s_{1234},m_t^2)-I_2(s_{12345},m_t^2)\right)
-\frac{1}{s_{12345}-s_{1234}}\bigg) \nonumber
\end{eqnarray}
Here $I_3(s,t,m_t^2)$ and $I_2(s,m_t^2)$ represent the triangle with two massive external legs and the bubble
integral, respectively; in both cases the internal propagators have a common mass, $m_t$.
In the notation of the QCDLoop library~\cite{Ellis:2007qk}, which we use to evaluate the integrals,
\begin{equation}
I_3(s,t,m_t^2) \equiv I_3(s,t,0; m_t^2,m_t^2,m_t^2) \quad
\mbox{and} \quad
I_2(s,m_t^2) \equiv I_2(s; m_t^2,m_t^2) \,.
\end{equation}

\bibliographystyle{JHEP}
\bibliography{VH}

\end{document}